\pdfoutput=1
\documentclass[letterpaper,titlepage,11pt]{article}

\usepackage{jheppub}
\usepackage[T1]{fontenc}
\usepackage[latin9]{inputenc}
\usepackage[active]{srcltx}
\usepackage{textcomp}
\usepackage{amsmath}
\usepackage{amssymb}
\usepackage{esint}
\usepackage{multirow}
\usepackage{empheq}
\usepackage{textcomp}
\usepackage{xcolor}

\def\cW{\mathcal{W}}

\usepackage{enumerate}
\usepackage{xcolor}
\def\be{\begin{eqnarray}}
\def\ee{\end{eqnarray}}
\def\beann{\begin{eqnarray*}}
\def\eeann{\end{eqnarray*}}
\def\beq{\begin{equation}}
\def\eeq{\end{equation}}
\def\ba{\begin{array}}
\def\ea{\end{array}}
\def\ben{\begin{enumerate}}
\def\een{\end{enumerate}}
\def\bea{\begin{eqnarray}}
\def\eea{\end{eqnarray}}

\def\nonu{\nonumber}
\def\pa{\partial}

\providecommand{\Lt}{{\tt L}}
\renewcommand{\Lt}{{\tt L}}
\providecommand{\Wt}{{\tt W}}
\renewcommand{\Wt}{{\tt W}}

\providecommand{\Gt}{{\tt G}}
\renewcommand{\Gt}{{\tt G}}

\providecommand{\At}{{\tt A}}
\renewcommand{\At}{{\tt A}}
\providecommand{\St}{{\tt S}}
\renewcommand{\St}{{\tt S}}

\providecommand{\Jt}{{\tt J}}
\renewcommand{\Jt}{{\tt J}}

\def\cW{{\cal W}}

\def\be{\begin{equation}}
\def\ee{\end{equation}}
\def\bea{\begin{eqnarray}}
\def\eea{\end{eqnarray}}
\def\ba{\begin{array}}
\def\ea{\end{array}}

\usepackage{amsfonts}

\usepackage{tikz}

\usepackage{bbm}

\usepackage{lineno}

\title{\bf{On the $\mathcal{N}=3$  and  $\mathcal{N}=4$ superconformal holographic dictionary}}
\newcommand{\itu}{\dagger}

\author[\itu]{H. T. \"Ozer}
\emailAdd{ozert@itu.edu.tr}
\author[\itu]{,\,\,\,Ayt\"ul Filiz}
\emailAdd{aytulfiliz@itu.edu.tr}

\affiliation[\itu]{Istanbul Technical University,\,Faculty of Science and Letters,
\,Physics Department,\\34469 Maslak,\,Istanbul,Turkey.}
\abstract{
This study presents comprehensive examples of $\mathfrak{osp}(\mathcal{N}|2)$ Chern\,-\,Simons supergravity on $AdS_3$ for $\mathcal{N}>2$. These formulations, which include the most general boundary conditions, represent extensions of previously discovered works  (\textit{Ozer and Filiz,Eur Phys J C 82(5):472, 2022}) for $\mathcal{N}<3$. In our work, we show that under the loosest set of boundary conditions, the asymptotic symmetry algebras consist of two copies of the $\mathfrak{osp}(3|2)_k$ and $\mathfrak{osp}(4|2)_k$ algebras. We subsequently restrict the gauge fields upon the boundary conditions to achieve supersymmetric extensions of the Brown\,-\,Henneaux boundary conditions. Based on these results, we finally find that the asymptotic symmetry algebras are two copies of the $\mathcal{N}=3$ and $\mathcal{N}=4$ superconformal algebras for $\mathcal{N}=(3,3)$ and $\mathcal{N}=(4,4)$ extended higher\,-\,spin supergravity theory in $AdS_3$.
}
\makeatother
\begin{document}
\today
\maketitle
\section{Introduction}
\hspace{0.5cm}
The holographic conjecture states that quantum gravity within a $(d+1)$\,-\,dimensional spacetime is equivalent to a quantum field theory at the boundary of the $d$\,-\,dimensional space. This conjecture is a powerful tool for studying quantum gravity. An important example of this conjecture is the holographic duality between $(d+1)$\,-\,dimensional Anti\,-\,de Sitter (AdS) gravity and conformal field theory in dimension $d$ \cite{Maldacena:1997re}. This duality, first proposed by Maldacena in 1997, provides a fundamental framework for studying the complex interplay between gravitational and quantum field theoretical phenomena, offering a comprehensive and insightful perspective on the nature of these fundamental forces.

Conformal Field Theory (CFT) is a fundamental concept in theoretical physics, with applications across diverse fields, from string theory to condensed matter theory. The strength of CFTs lies in their symmetries, which impose significant constraints on the theory's structure and content. The conformal symmetry is particularly enhanced in two dimensions, resulting in an infinite set of generators known as the Virasoro algebra \cite{Virasoro:1969zu}. Researchers have explored various extensions of the two\,-\,dimensional conformal algebra beyond the Virasoro algebra, including the Kac\,-\,Moody algebra \cite{Kac_1968,Moody:1968zz} and the W\,-\,algebra \cite{Zamolodchikov:1985wn}, as well as the Virasoro algebra itself \cite{DiFrancesco:1997nk,Blumenhagen:2009zz,Prochazka:2014gqa,Ozer:2015dha}.

However, extending the Virasoro algebra to its supersymmetric counterpart, the super Virasoro algebra presents increasing challenges. Despite extensive investigations into supersymmetric extensions with a limited number of supersymmetries ($N\leq 4$) \cite{Ademollo:1975an,Ademollo:1976pp,Schoutens:1987uu,Schoutens:1988ig,Sevrin:1988ew,Eguchi:1987sm,Eguchi:1987wf,Odake:1988bh,Odake:1989dm}, complete construction of the super Virasoro algebra remains elusive for $N>4$. The complexities involved in this endeavor highlight the intricate nature of supersymmetric extensions and underscore the need for further exploration into the unexplored domain of higher supersymmetry orders.
The Hamiltonian reduction of affine algebras gives rise to higher spin algebras. These can be described through flat connections in the Drinfeld\,-\,Sokolov form, as extensively documented in \cite{Drinfeld:1984qv}. This characterization aligns seamlessly with the overarching framework of higher spin dualities. These dualities specifically involve three\,-\,dimensional Chern\,-\,Simons theory and two\,-\,dimensional conformal field theories (CFTs) featuring $\mathcal{W}$\,-\,symmetry. Noteworthy instances include the well\,-\,established correlation between $\mathfrak{sl}(N)$ Chern\,-\,Simons theory and CFTs endowed with $\mathcal{W}_N$\,-\,symmetry \cite{Campoleoni:2010zq,Campoleoni:2011hg}. Furthermore, recent and significant examples include the duality observed between $\mathfrak{hs}(\lambda)$ Chern\,-\,Simons theory and CFTs characterized by $\mathcal{W_\infty}(\lambda)$\,-\,symmetry \cite{Henneaux:2010xg,Gaberdiel:2011wb}. A comprehensive review of this subject is available in \cite{Gaberdiel:2012uj}.

In their seminal work, Brown and Henneaux presented a groundbreaking analysis \cite{Brown:1986nw}, revealing that the asymptotic symmetries of $\mathrm{AdS_3}$ gravity manifest as two copies of the Virasoro algebra. This discovery serves as a precursor to the $\mathrm{AdS/CFT}$ correspondence \cite{Maldacena:1997re} and embodies the Holographic Principle \cite{Susskind:1994vu}. Subsequent advancements in $\mathrm{AdS_3}$ holography have expanded this foundation. Noteworthy extensions include investigations into the asymptotic symmetries of supergravity and extended supergravity, as documented in \cite{Banados:1998pi} and \cite{Henneaux:1999ib}, respectively. These studies conclusively established that the asymptotic symmetry algebra for (extended) supergravity aligns with the (extended) superconformal algebra \cite{Goddard:1986ee, Schwimmer:1986mf}.

The exploration of diverse three\,-\,dimensional asymptotic symmetries within supergravity models remains a subject of considerable interest, as it can provide insights into the geometric origins of superconformal algebras. Noteworthy instances in this line of investigation include the asymptotic analysis of supersymmetric higher\,-\,spin gravity \cite{Henneaux:2012ny,Tan:2012xi,Gaberdiel:2014yla}, hypergravity \cite{Henneaux:2015ywa,Fuentealba:2015wza}, and supergravity in the at\,-\,space limit \cite{Barnich:2014cwa,Lodato:2016alv,Banerjee:2017gzj,Fuentealba:2017fck,Basu:2017aqn}, as well as \cite{Banerjee:2018hbl}. The Chern\,-\,Simons formulation of three\,-\,dimensional gravity, introduced by Witten \cite{Witten:1988hc}, emerges as a crucial tool in deriving these results. This formulation proves advantageous for a more straightforward analysis of these symmetries than the metric formulation, as elucidated in pedagogical reviews such as \cite{Banados:1994tn} and \cite{Banados:1998sm}.

Selecting of fall\,-\,off conditions for metric parameters is essential in the $\mathrm{AdS_3}$ holography. Among the widely employed conditions for pure $\mathrm{AdS_3}$ gravity are the Brown\,-\,Henneaux boundary conditions \cite{Brown:1986nw}. Despite their prevalence, alternative boundary conditions have been introduced in recent years. Notably, Compere, Song, and Strominger \cite{Compere:2013bya}, as well as Troessaert \cite{Troessaert:2013fma}, have proposed novel sets, resulting in the emergence of two instances of a warped conformal algebra with distinct Kac\,-\,Moody levels. Avery, Poojary, and Suryanarayana \cite{Avery:2013dja} introduced boundary conditions that yield a semidirect sum of an $\mathfrak{sl}(2)_k$ algebra and a Virasoro algebra. In their comprehensive study, Grumiller and Riegler \cite{Grumiller:2016pqb} investigated the most general boundary conditions for $\mathrm{AdS_3}$ gravity. Consequently, the asymptotic symmetry algebra was found to be generated by two copies of the $\mathfrak{sl}(2)_k$ algebra. Furthermore, it is noteworthy that the previously identified boundary conditions were derived by imposing certain constraints on these most general conditions.

This methodology has been applied across various scenarios, including three\,-\,dimensional AdS space \cite{Grumiller:2017sjh}, chiral higher spin gravity \cite{Krishnan:2017xct, Ozer:2017dwk}, and (supersymmetric) two\,-\,dimensional gravity \cite{Grumiller:2017qao, Cardenas:2018krd}. This paper endeavors to broaden the scope of the analysis initiated in \cite{Grumiller:2016pqb} to three\,-\,dimensional AdS supergravity. We systematically derive the asymptotic symmetry algebra for the most permissive set of (extended) supergravity boundary conditions.\,We reaffirmed some prior findings while presenting novel boundary conditions linked to warped superconformal algebras. Consequently, this methodology is an invaluable platform for delving into the intricate asymptotic structure of (extended) supergravity.

This paper addresses a hitherto unresolved phenomenon by presenting two candidate solutions for the most general $\mathcal{N}=(3,3)$ and $\mathcal{N}=(4,4)$ extended higher spin supergravity theories on $AdS_3$. Building upon our previous studies \cite{Ozer:2019nkv,Ozer:2021wtx}, it is crucial to underscore that our inspiration from Grumiller's seminal work \cite{Grumiller:2016pqb} continues to guide this investigation. The primary objective of this work is to determine the explicit form of the holographic dictionary between $\mathfrak{osp}(\mathcal{N}|2)$ Chern\,-\,Simons supergravity on $AdS_3$ and two\,-\,dimensional $CFT_2$ with $\mathcal{N}=3$ and $\mathcal{N}=4$ superconformal symmetry, respectively. Demonstrating that our theory aligns with the metric class established by \cite{Grumiller:2017sjh}, we observe that the metric formulation can encompass both charge and chemical potentials present in the Chern\,-\,Simons formalism. This offers an alternative solution to the non\,-\,chiral Drinfeld\,-\,Sokolov type boundary conditions.

Our primary focus is on the extended $\mathcal{N}=(3,3)$ and $\mathcal{N}=(4,4)$ Chern\,-\,Simons theory based on $\mathfrak{osp}(3|2)$ and $\mathfrak{osp}(4|2)$ superalgebras, respectively. This results in an asymptotic symmetry algebra comprising two copies of the $\mathfrak{osp}(3|2)_k$ and $\mathfrak{osp}(4|2)_k$ affine algebras. We impose specific restrictions on the gauge fields under the most general boundary conditions, leading to supersymmetric extensions of the Brown\,-\,Henneaux Boundary conditions. Furthermore, we establish that the asymptotic symmetry algebras reduce to two copies of the $\mathcal{N}=3$ and $\mathcal{N}=4$ superconformal algebras in $AdS_3$ supergravity theory.

It is noteworthy that exploring a distinct class of boundary conditions for (super)gravity, as emerging in the literature (e.g., \cite{Compere:2013bya,Troessaert:2013fma,Avery:2013dja}), presents an intriguing problem. Their higher spin generalization is not as evident as the boundary conditions proposed by Grumiller and Riegler. In light of these findings, it is clear that this method provides an excellent laboratory for investigating the rich asymptotic structure of extended higher\,-\,spin supergravity.

This paper is structured as follows: To establish a foundation, we begin with a concise overview of the fundamental aspects of the $\mathcal{N}=3$ and $\mathcal{N}=4$ superconformal algebras in the initial sections. Subsequently, in Section 3, we present a comprehensive formulation of $\mathfrak{osp}(\mathcal{N}|2) \oplus \mathfrak{osp}(\mathcal{N}|2)$ Chern\,-\,Simons supergravity. Following a brief examination of $\mathfrak{osp}(\mathcal{N}|2) \oplus \mathfrak{osp}(\mathcal{N}|2)$ Chern\,-\,Simons supergravity on $AdS_3$, Sections 3 and 4 are dedicated to establishing a clear and explicit holographic correspondence between $\mathfrak{osp}(\mathcal{N}|2)$ Chern\,-\,Simons supergravity on $AdS_3$ and two\,-\,dimensional $CFT_2$ possessing $\mathcal{N}=3$ and $\mathcal{N}=4$ superconformal symmetry. This analysis extends to encompass both affine and superconformal boundaries. Additionally, we explore the emergence of asymptotic symmetry and higher spin Ward identities from the bulk equations of motion coupled to spin currents.

In these sections, we examine the characterization of asymptotic symmetry algebras as classical duplicates of the $\mathfrak{osp}(\mathcal{N}|2)_k$ and affine algebra on the affine boundary. Simultaneously, we establish the $\mathfrak{so}(\mathcal{N})$ superconformal symmetry algebra for $\mathcal{N}=3$ and $\mathcal{N}=4$ on the superconformal boundary. Furthermore, we elaborate on the description of chemical potentials linked to source fields, manifesting through the temporal components of the connection.

In the concluding section, we summarize our findings and insights,\,address any unresolved questions, and propose potential avenues for future research.

\section{The superconformal algebras:}\label{sec2}

We will start by going over the basics of the ${\mathcal N}=3$ superconformal algebra.\,This section is not meant to be a comprehensive  survey;\,instead,\,it serves as a summary of
the key features relevant to our  objectives.\,The $\mathcal{N}=3$ superconformal algebra  comes in two guises  as a linear \cite{Ademollo:1975an,Ademollo:1976pp} and a non-linear
\cite{Knizhnik:1986wc,Bershadsky:1986ms} versions in the literature.\,we are first interested in the linear version version of the algebra.
\subsection{The $\mathcal{N}=3$ linear\,-\,superconformal algebras:}\label{sec21}

The 8 currents are given by a single spin-$\frac{1}{2}$ current $\Psi(z)$, three spin-1 currents $J^a(z)$, $a=1,2,3$ transforming as a vector representation under $\mathfrak{so}(3)$,
three $\frac{3}{2}$ currents $G^i(z)$, $i=1,2,3$ transforming as a vector representation under $\mathfrak{so}(3)$,\,and the spin-2 current $T(z)$.\,In particular, the spin-$\frac{1}{2}$
 and spin-2 currents are $\mathfrak{so}(3)$ singlets, all having the standard mode expansions
{
$
T(z)=  \sum_n L_n z^{-n-2}~,\,
G^i(z)= \sum_r G^i_r z^{-r-3/2}~,\,
J^a(z)= \sum_n J^a_n z^{-n-1}~,\
$ and  $
F(z)= \sum_r F_r z^{-r-1/2}~
$.\\

Non-zero of the $OPEs$ defining the $\mathcal{N}=3$ superconformal algebra take the standard linear form,
{\bea
F(z_1) \, F(z_2) & \sim & \frac{1}{z_{12}} \, \frac{c}{3} , \\
F(z_1) \, J^i(z_2) & \sim & 0 ,
  \\
F(z_1) \, G^i(z_2) & \sim & \frac{1}{z_{12}} \, J^i  ,  \\
J^i(z_1) \, J^j(z_2) & \sim & \frac{1}{z_{12}^2} \, \frac{c}{3} \delta^{ij} +
\frac{1}{z_{12}} \, i \epsilon^{ijk} J^k  ,  \\
 J^i(z_1) \, G^j(z_2) & \sim & \frac{1}{z_{12}^2} \,  \delta^{ij} F  +
\frac{1}{z_{12}} \, i \epsilon^{ijk} G^k  ,  \\
G^i(z_1) \, G^j(z_2) & \sim & \frac{1}{z_{12}^3} \, \frac{2c}{3} \delta^{ij}
+ \frac{1}{z_{12}^2} \, 2 i \epsilon^{ijk} J^k \nonumber\\ 
&+& \frac{1}{z_{12}}
\left( 2 \delta^{ij} T + i \epsilon^{ijk} \partial J^k \right)  , \\
T(z_1) \, F(z_2) & \sim  &
\frac{1}{z_{12}^2} \, \frac{1}{2} F + \frac{1}{z_{12}} \partial F, \\
T(z_1) \, J^i(z_2) & \sim  &
\frac{1}{z_{12}^2} \,  J^i  + \frac{1}{z_{12}} \partial J^i  ,\\
%
T(z_1) \, G^i(z_2) & \sim  &
\frac{1}{z_{12}^2} \, \frac{3}{2} G^i  + \frac{1}{z_{12}} \partial G^i  ,  \\
T(z_1) \, T(z_2) & \sim & \frac{1}{z_{12}^4} \, \frac{c}{2} +
\frac{1}{z_{12}^2} \, 2 T  + \frac{1}{z_{12}} \partial T.
\label{comfund2}
\eea}

where $z_{12}=z_1-z_2$}, $c=3k$ is the central charge and $k$ is an arbitrary `level' of the Kac\,-\,Moody subalgebra, $\epsilon^{ijk}$ are the structure constants of $\mathfrak{so}(3)$ in the fundamental (vector) representation,

The nonlinear version of above ${\mathcal N}=3$ superconformal algebra can be obtained by factoring out the spin-$\frac{1}{2}$ current \cite{Knizhnik:1986wc,Bershadsky:1986ms} and becomes the result of $(\ref{qalg11} )$-$(\ref{qalg12})$.
\subsection{ The $ {\mathcal N}=3$ nonlinear\,-\,Knizhnik Bershadsky algebra }

We have only looked at the ${\mathcal N}=3$ linear superconformal algebra  so far.\,Among the eight currents of the ${\mathcal N}=3$ linear  superconformal algebra,\,We can decouple the spin-$\frac{1}{2}$
current $F(z)$  from the other seven remaining currents by doing something like \cite{Ahn:2013oya,Ahn:2014via,Ahn:2015gba}.\,One get the eight higher spin currents and their OPEs after decoupling the above
spin-$\frac{1}{2}$ current in this section.\\

The spin-$1$ currents of the ${\mathcal N}=3$ superconformal algebra do not have any singular terms with the spin-$\frac{1}{2}$  current.\,Then it follows that the  spin-$1$ currents should be used as the previous
one $J^i(z)$.\,We should acquire the new spin-$\frac{3}{2}$  currents and  spin-$2$ currents for the spin-$\frac{3}{2}$ currents and spin-$2$ currents.\,The new seven currents are really specified as \cite{GS,Schoutens}.\\
\bea
\hat{T}(z) & = & T(z) + \frac{3}{2c} F \pa F(z),   \\
\hat{G}^i(z)  & = & G^i(z) -\frac{3}{c} J^i F(z),   \\
\hat{J}^i(z) & = & J^i(z).
\label{tgj}
\eea
The relative coefficients on the right hand side of (\ref{tgj}) can be set by requiring that the requirements:
$
F(z_1) \; \hat{T}(z_2)    \sim   0, \,
F(z_1) \; \hat{G}^i(z_2)  \sim 0,  \,
$ and $
F(z_1) \; \hat{J}^i(z_2)  \sim  0
$
be met.
Then, using (\ref{tgj}), we can compute the OPEs between the new seven currents,\,which may be summed as \cite{Knizhnik:1986wc,Bershadsky:1986ms}
{\small
\bea
\hat{J}^i(z_1) \, \hat{J}^j(z_2) & \sim & \frac{1}{z_{12}^2} \,
\frac{c}{3} \delta^{ij} +
\frac{1}{z_{12}} \, i \epsilon^{ijk} \hat{J}^k \label{qalg11} ,
 \\
 \hat{J}^i(z_1) \, \hat{G}^j(z_2) & \sim &
\frac{1}{z_{12}} \, i \epsilon^{ijk} \hat{G}^k  ,
  \\
\hat{G}^i(z_1) \, \hat{G}^j(z_1)  & \sim &
\frac{1}{z_{12}^3} \, \frac{1}{3} (2 c-3) \delta^{ij}
+ \frac{1}{z_{12}^2} \,  \frac{i (2 c-3) }{c}
\epsilon^{ijk} \hat{J}^k \nonu  \\
& + &  \frac{1}{z_{12}}
\bigg( 2 \delta^{ij} \hat{T} + i \epsilon^{ijk} \pa \hat{J}^k
-\frac{3}{c} \hat{J}^i \hat{J}^j
\bigg) ,\\
%
\hat{T}(z_1) \, \hat{J}^i(z_2) & \sim &
\frac{1}{z_{12}^2} \,  \hat{J}^i  + \frac{1}{z_{12}} \pa \hat{J}^i  ,
  \\
\hat{T}(z_1) \, \hat{G}^i(z_2) & \sim &
\frac{1}{z_{12}^2} \, \frac{3}{2} \hat{G}^i  +
\frac{1}{z_{12}} \pa \hat{G}^i  ,
  \\
\hat{T}(z_1) \, \hat{T}(z_2) & \sim & \frac{1}{z_{12}^4} \, \frac{1}{4} (2 c-1) +
\frac{1}{z_{12}^2} \, 2 \hat{T}  + \frac{1}{z_{12}} \pa \hat{T}  .
\label{qalg12}
\eea}

It is worth noting that the OPE has a nonlinear term between the spin-$\frac{3}{2}$ currents.\,The aforementioned $OPEs$ $(\ref{qalg11} )$-$(\ref{qalg12})$ easily become the one in \cite{Knizhnik:1986wc}
(where there is an error in the $OPE$) if one changes $i \hat{J}^i(z) =J_K^i(z)$  and other currents remain unaltered.\\
\subsection{The $ \mathcal{N}=4 $ linear\,-\,superconformal algebra }\label{sec2}

The $\mathcal{N}=4$ algebra was first given by Ademollo at al.\cite{Ademollo:1976wv},\,This algebra does not contain the $\mathcal{N}=3$  algebra discussed above.\,They both contain an
$\mathfrak{su}(2)$ Kac-moody algebra but the supercharges transform as an $\mathfrak{su}(2)$ triplet in the $\mathcal{N}=3$  algebra and as a doublet in the $\mathcal{N}=4$ algebra.\,
The $ \mathcal{N}=4 $  superconformal algebra algebra contains a Virasoro sub-algebra, and $\mathfrak{su}(2)$ Kac-Moody algebra.\,The four super charges form a complex $\mathfrak{su}(2)$
doublet.\,Therefore,\,the 8 currents are given by  three spin-1 currents $J^i(z)$, $i=1,2,3$ transforming as a complex $\mathfrak{su}(2)$ doublet, four $\frac{3}{2}$ currents
$G^a(z)$ and $\bar{G}^a(z)$, $a=1,2$ transforming as an $\mathfrak{su}(2)$ doublets,\,all having the standard mode expansions
{
$
T(z)=  \sum_n L_n z^{-n-2}~,\,
G^a(z)= \sum_r G^a_r z^{-r-3/2}~,\,
\bar{G}^a(z)= \sum_r \bar{G}^a_r z^{-r-3/2}~,\,
$ and  $
J^i(z)= \sum_n J^i_n z^{-n-1}
$
.\\

Non-zero of the  $OPEs$ defining the $\mathcal{N}=4$ superconformal algebra take the standard linear form,
{\bea
 J^i(z_1) \, J^j(z_2) & \sim & \frac{1}{z_{12}^2} \, \frac{k}{2} \delta^{ij} +\frac{1}{z_{12}} \, i \epsilon^{ijk} J^k  ,  \\
 J^i(z_1) \, G^a(z_2) & \sim &  -\frac{1}{z_{12}} \, \frac{1}{2} (\sigma^{i})_{ab} G^b  ,  \\
  J^i(z_1) \, \bar{G}^a(z_2) & \sim & \frac{1}{z_{12}} \, \frac{1}{2} (\sigma^{i})^{*}_{ab} \bar{G}^b  ,  \\
  G^a(z_1) \, \bar{G}^b(z_2) & \sim & \frac{1}{z_{12}^3} \, 4k \delta^{ab}\
  + \frac{1}{z_{12}^2} \,  4(\sigma^{i})_{ab} J^i \nonumber\\
  &+& \frac{2}{z_{12}}\left(  \delta^{ab} T
  +  (\sigma^{i})_{ab} \partial J^i \right)  , \\
   T(z_1) \, J^i(z_2) & \sim & \frac{1}{z_{12}^2} \,  J^i  + \frac{1}{z_{12}} \partial J^i  ,\\
   T(z_1) \, G^a(z_2) & \sim & \frac{1}{z_{12}^2} \, \frac{3}{2} G^a  + \frac{1}{z_{12}} \partial G^a  ,  \\
   T(z_1) \, \bar{G}^a(z_2) & \sim & \frac{1}{z_{12}^2} \, \frac{3}{2} \bar{G}^a  + \frac{1}{z_{12}} \partial \bar{G}^a  ,  \\
   T(z_1) \, T(z_2)   & \sim & \frac{1}{z_{12}^4} \, \frac{c}{2} +\frac{1}{z_{12}^2} \, 2 T  + \frac{1}{z_{12}} \partial T.
\label{comfund1}
\eea}
where $c=6k$ is the central charge and $k$ is an arbitrary `level' of the Kac\,-\,Moody subalgebra, $\epsilon^{ijk}$ are  $\mathfrak{su}(2)$ structure constants and $\sigma^{i}$ are the Pauli matrices.
\section{The $so(\mathcal{N})$ Holographic Dictionary}\label{sec21}

After examining some properties of the $\mathcal{N}=3$ and $\mathcal{N}=4$ superconformal algebras, we now examine into their realization in terms of higher spin fields on $AdS_3$. This exploration begins with a concise overview of Chern\,-\,Simons supergravity theory, focusing specifically on aspects relevant to constructing the holographic dictionary. Subsequently, we present a detailed derivation of the asymptotic symmetry algebra for the $so(\mathcal{N})$ case. Following this derivation, we engage in a discussion of sources and the corresponding holographic Ward identities. These identities play a crucial role in the examination of higher spin black hole solutions.

\subsection{Chern-Simons Supergravity in Three Dimensions}\label{sec2}

This section provides a concise overview of $AdS_3$ higher spin supergravity employing the Chern-Simons formalism. We specifically employ this formalism to investigate $AdS_3$ supergravity in the context of the $\mathfrak{osp}(\mathcal{N}|2)$ superalgebra basis.

\subsection{Connection to $\mathfrak{osp}(\mathcal{N}|2) \oplus \mathfrak{osp}(\mathcal{N}|2)$ Chern-Simons Supergravity}

In three dimensions, the Einstein-Hilbert action for supergravity with a negative cosmological constant can be equivalently formulated using Chern-Simons theory over a spacetime manifold $\mathcal{M}$:
\begin{equation}
    S = S_{CS}[\Gamma] - S_{CS}[\bar{\Gamma}],
\end{equation}
where
\begin{equation}
    S_{CS}[\Gamma] = \frac{k}{4\pi}\int_{\mathcal{M}} \mathfrak{str}\bigg(\Gamma\wedge \mathrm{d}\Gamma + \frac{2}{3}\, \Gamma\wedge\Gamma\wedge\Gamma \bigg),
\end{equation}
first introduced by Achucarro and Townsend \cite{Achucarro:1986uwr} and further developed by Witten \cite{Witten:1988hc}.

The Chern-Simons level $k$ is eventually related to the ratio of the $AdS_3$ radius $l$ and Newton's constant $G$, as well as the central charge $c$ of the superconformal field theory.

It is important to note that the 1-forms $(\Gamma, \bar{\Gamma})$ take values in the gauge group of the $\mathfrak{osp}(\mathcal{N}|2)$ superalgebra. The supertrace $\mathfrak{str}$, representing a metric on the $\mathfrak{osp}(\mathcal{N}|2)$ Lie superalgebra, is taken over the superalgebra generators. The $\mathfrak{osp}(\mathcal{N}|2)$ superalgebra is defined as the even and odd parts, respectively. To describe (extended) supergravity, one employs the even part as $\mathfrak{sl}(2) \oplus \mathfrak{so}(\mathcal{N})$ with generators $L_I$ ($I=0,\pm1$) and $T_{AB}=-T_{BA}$ ($A,B=1,2,...,\mathcal{N}$). The odd part $\mathfrak{g}_1$ has generators $G^\alpha_A$ ($\alpha=\pm\frac{1}{2}$). These generators satisfy the following algebraic relations:
\begin{eqnarray}
    \left[L_I, L_J\right] &=& (I-J) L_{I+J}, \\
    \left[L_{I}, G^A_{\alpha} \right] &=& \left(\frac{I}{2} - \alpha\right) G^A_{I+\alpha}, \\
    \left[ G^A_\alpha, T^{AB} \right] &=& \delta^{AB} G^C_\alpha - \delta^{AC} G^B_\alpha, \\
    \left[ T^{AB}, T^{CD} \right] &=& \delta^{AD} T^{BC} + \delta^{BC} T^{AD}  \nonumber \\
                                  &-& \delta^{AC} T^{BD} - \delta^{BD} T^{AC}, \\
    \left\{ G^A_{\alpha}, G^B_{\beta} \right\} &=& -2 L_{\alpha+\beta} \delta^{AB} - \left(\alpha-\beta\right) T^{AB}.
\end{eqnarray}

Here, we present some of the most notable relations.

\begin{itemize}
\item $\mathfrak{osp}(1|2)$: \\
For $\mathcal{N} = 1$, the generators $T^{AB}$ are no longer available, and one can drop the $\mathfrak{so}(\mathcal{N})$ indices. Therefore, the $\mathfrak{osp}(1|2)$ algebra is given by:
\begin{align}
        [L_I, L_J] &= (I - J) L_{I+J}, \quad [L_I, G_\alpha] = \left(\frac{I}{2} - \alpha\right)G_{I+\alpha}, \quad \{ G_\alpha, G_\beta \} = -2L_{\alpha+\beta}. \label{eq:osp12-algebra}
    \end{align}
    \item $\mathfrak{osp}(2|2)$: \\
    For $\mathcal{N} = 2$, due to the antisymmetry of the $\mathfrak{so}(2)$ generators, we can write $T^{AB} = \epsilon^{AB}\Jt$. Therefore, the algebra reduces to the $\mathfrak{osp}(2|2)$ superalgebra. We denote the bosonic generators as $\Jt$, $\Lt_I$ $(I=\pm1,0)$, and the fermionic ones as $\Gt_\alpha^A$ $(\alpha=\pm\frac{1}{2}, A=1,2)$. Their commutation relations read:
    \begin{align}
        [L_I, L_J] &= (I - J)L_{I+J}, \quad [L_I, G_\alpha^A] = \left(\frac{I}{2} - \alpha\right)G_{I+\alpha}^A, \label{eq:osp22-algebra-a} \\
        [G_\alpha^A, \Jt] &= \epsilon^{AB}G_\alpha^B, \quad \{ G_\alpha^A, G_\beta^B \} = -2L_{\alpha+\beta}\delta^{AB} - (\alpha - \beta)\epsilon^{AB}\Jt. \label{eq:osp22-algebra-b}
    \end{align}
\item $\mathfrak{osp}(3|2)$:\\
For $\mathcal{N} = 3$, owing to the antisymmetry of the $\mathfrak{so}(3)$ generators, we can express $T^{ABC} = \epsilon^{ABC}\Jt^C$. Consequently, the algebra reduces to the $\mathfrak{osp}(3|2)$ superalgebra.
We denote the bosonic generators by $\Jt^A$ ($A=1,2,3$), $\Lt_{I}$ ($I=\pm1,0$), and the fermionic generators by $\Gt_{\alpha}^{A}$ ($\alpha=\pm\frac{1}{2}$, $A=1,2,3$). Their commutation relations are given by
\begin{align}
\label{algebraLG}
\left[\Lt_{I},\Lt_{J}\right] &= \left(I-J\right)\Lt_{I+J},\\
\left[\Lt_{I},\Gt_{\alpha}^{A}\right] &= \left(\frac{I}{2}-\alpha\right)\Gt_{I+\alpha}^{A},\\
\left[\Jt^A,\Jt^B\right] &= \epsilon^{ABC}\Jt^C,\\
\left[\Jt^A,\Gt_{\alpha}^{B}\right] &= \epsilon^{ABC}\Gt_{\alpha}^{C},\\
\left\{\Gt_{\alpha}^{A},\Gt_{\beta}^{B}\right\} &= 2\,\delta^{AB}\Lt_{\alpha+\beta}\nonumber\\
             &-\left(\alpha-\beta\right)\delta_{\alpha,-\beta}\epsilon^{ABC} \Jt^C,
\end{align}
except for zero commutators.\\
It is important to note that there are variants with different spin contents, referred to as the $\mathfrak{osp}(3|2)$ superalgebra in the literature \cite{Ozer:2019nkv}. Specifically, a copy of the $\mathcal{N}=2$ superalgebra arises when the even-graded sector of the superalgebra is decomposed into a spin-2 component, the $\mathfrak{sl}(2)$ generators $\Lt_{i}$ ($i=\pm1,0$), a spin-1 element $\Jt$, a spin-2 multiplet $\At_{i}$ ($i=\pm1,0$), and a spin-3 multiplet $\Wt_{i}$ ($i=\pm2,\pm1,0$). All these generators together span the bosonic sub-algebra $\mathfrak{sl}(3) \oplus \mathfrak{sl}(2) \oplus \mathfrak{u}(1)$. Furthermore, the odd-graded elements decompose into two spin-$\frac{3}{2}$ multiplets $\Gt_{r}^{M}$ ($r=\pm\frac{1}{2}$, $M=\pm$) and two spin-$\frac{5}{2}$ multiplets $\St_{r}^{M}$ ($r=\pm\frac{3}{2},\pm\frac{1}{2}$, $M=\pm$).

\item $\mathfrak{osp}(4|2)$: \\
For the case $\mathcal{N} = 4$ where $\mathfrak{so}(4) \simeq \mathfrak{su}(2) \oplus \mathfrak{su}(2)$, special treatment is required. We denote the bosonic generators as $\Jt^A$ $(A=1,2,3)$, $\Lt_{I}$ $(I=\pm1,0)$, and the fermionic generators as $\Gt_{\alpha}^{a}$, $\bar{\Gt}_{\alpha}^{a}$ $(\alpha=\pm\frac{1}{2},\,a=1,2)$. Their commutation relations are given by
\begin{align}
    \label{algebraLG}
    [\Lt_{I},\Lt_{J}] & = (I-J)\Lt_{I+J}, \\
    [\Lt_{I},\Gt_{\alpha}^{a}] & = \left(\frac{I}{2}-\alpha\right)\Gt_{I+\alpha}^{a}, \\
    [\Lt_{I},\bar{\Gt}_{\alpha}^{a}] & = \left(\frac{I}{2}-\alpha\right)\bar{\Gt}_{I+\alpha}^{a}, \\
    [\Jt^A,\Jt^B] & = \epsilon^{ABC}\Jt^C, \\
    [\Jt^A,\Gt_{\alpha}^{a}] & = (\sigma^{A})_{ab}\Gt_{\alpha}^{b}, \\
    [\Jt^A,\bar{\Gt}_{\alpha}^{a}] & = (\sigma^{A})^{*}_{ab}\bar{\Gt}_{\alpha}^{b}, \\
    \{\Gt_{\alpha}^{a},\bar{\Gt}_{\beta}^{b}\} & = 2\delta^{ab}\Lt_{\alpha+\beta}\nonumber\\ 
    &- (\alpha-\beta)\delta_{\alpha,-\beta}(\sigma^{c})_{ab} \Jt^c,
\end{align}
except for zero commutators. Here, $\epsilon^{ijk}$ represents the structure constants of $\mathfrak{su}(2)$, and $\sigma^{i}$ denotes the Pauli matrices.
\end{itemize}

The equations of motion for Chern-Simons theory, often referred to as the flatness conditions, correspond to the vanishing of the field strengths, i.e., \( F = \bar{F} = 0 \), where
\begin{equation}
\label{flatness}
    F = \mathrm{d}\Gamma + \Gamma\wedge \Gamma = 0, \quad \bar{F} = \mathrm{d}\bar{\Gamma} + \bar{\Gamma}\wedge \bar{\Gamma} = 0.
\end{equation}
This condition is equivalent to Einstein's equations. The connection to Einstein's equations is established by expressing Lie algebra-valued generalizations of the vielbein and spin connection in terms of the gauge connections. Consequently, the metric \( g_{\mu \nu} \) can be obtained from the vielbein \( e = \frac{\ell}{2}(\Gamma - \bar{\Gamma}) \) as follows:
\begin{equation}
\label{metric}
    g_{\mu \nu} = \frac{1}{2}\,\mathfrak{str}( e_{\mu} e_{\nu} ).
\end{equation}

By adopting the radial gauge, asymptotically \( AdS_3 \) connections can be represented as
\begin{eqnarray}
\label{ads31}
    \Gamma &=& b^{-1} a\left(t, \phi\right) b + b^{-1} \mathrm{d}b, \quad
    \bar{\Gamma} = b \bar{a}\left(t, \phi\right) b^{-1} + b \mathrm{d}b^{-1},
\end{eqnarray}
where the state-independent group element \( b(\rho) \) is given by
\begin{eqnarray}
\label{gr}
    b(\rho) = e^{\Lt_{-1}} e^{\rho \Lt_{0}},
\end{eqnarray}
enabling a more general metric that encompasses all \( \mathfrak{osp}(\mathcal{N}|2) \) charges. It is crucial to note that, as long as \( \delta b = 0 \), the choice of \( b \) is irrelevant for asymptotic symmetries. This freedom allows for a more general metric, and it is essential to choose boundary conditions that preserve this generality for supergravity.

Furthermore, in the radial gauge, the connections \( a\left(t, \phi\right) \) and \( \bar{a}\left(t, \phi\right) \) are \(\mathfrak{osp}(3|2)\) Lie superalgebra-valued fields that are independent of the radial coordinate:
\begin{eqnarray}
\label{conn}
    a\left(t, \varphi\right) &=& a_{t}\left(t, \varphi\right)\mathrm{d}t + a_{\varphi}\left(t, \varphi\right)\mathrm{d}\varphi, \\
    \bar{a}\left(t, \varphi\right) &=& \bar{a}_{t}\left(t, \varphi\right)\mathrm{d}t + \bar{a}_{\varphi}\left(t, \varphi\right)\mathrm{d}\varphi.
\end{eqnarray}

Hereafter, we focus only on the unbarred sector since the analysis of the barred sector works in complete analogy, yielding the same outcomes. The same algorithm can be applied to the barred sector due to the procedure used.

\section{$\mathcal{N}=(3,3)$ $\mathfrak{osp}(3|2) \oplus \mathfrak{osp}(3|2)$ Higher-Spin Chern-Simons Supergravity}

We initiate the study of asymptotically $AdS_3$ boundary conditions for $\mathfrak{osp}(3|2) \oplus \mathfrak{osp}(3|2)$ Chern-Simons theory in the affine case. We illustrate how the procedure outlined in \cite{Grumiller:2016pqb} can be employed to assess the asymptotic symmetry algebra. Based on the obtained results, the most general solution of Einstein's equations that is asymptotically $AdS_3$ can be expressed using the following general metric form:

\begin{eqnarray}\label{sugra08}
    \mathrm{d}s^2 &=& \mathrm{d}\rho^2 + 2\left[ e^\rho N^{(0)}_i +  N^{(1)}_i +  e^{-\rho} N^{(2)}_i
    + \mathcal O \left( e^{-2\rho}\right)\right]\mathrm{d}\rho \mathrm{d}x^i\nonumber\\
    &+& \left[ e^{2\rho} g^{(0)}_{ij} +  e^\rho g^{(1)}_{ij} +  g^{(2)}_{ij}
    + \mathcal O \left( e^{-\rho}\right)\right]\mathrm{d}x^i \mathrm{d}x^j.
\end{eqnarray}

Therefore, it is crucial to define the most general $\mathcal{N}=(3,3)$ supergravity boundary conditions that preserve this form of the metric.

\subsection{Affine Boundary Conditions}\label{osp21}

The objective of this section is to construct $\mathcal{N}=(3,3)$ extended higher-spin $AdS_3$ supergravity as $\mathfrak{osp}(3|2)\oplus\mathfrak{osp}(3|2)$ Chern-Simons gauge theory on the affine boundary. We proceed with our calculations to elucidate the asymptotic symmetry algebra for the loosest set of boundary conditions. Therefore, we consider the principal embedding of $\mathfrak{osp}(1|2)$ into $\mathfrak{osp}(3|2)$ as a sub-algebra, giving rise to the asymptotic symmetry.

We start by proposing the $\mathfrak{osp}(3|2)$ Lie superalgebra valued $a_\varphi$ component of the gauge connection in the form:
\begin{eqnarray}\label{bouncond999}
a_\varphi &=&\rho_A\mathcal{J}^A\Jt^A +\gamma_i\mathcal{L}^i \Lt_{i}+\sigma_M^p\mathcal{G}_{A}^{p}\Gt_{p}^{A}
\end{eqnarray}

where $\rho_A=\frac{2}{k}$, $\gamma_{0}=-2\gamma_{\pm1}=-\frac{4}{k}$, and $\sigma_{A}^{-\frac{1}{2}}=-\sigma_{A}^{\frac{1}{2}}=-\frac{1}{k}$ are scaling parameters to be identified later. We have twelve state-dependent functions consisting of six bosonic $\left(\mathcal{J}^A,\mathcal{L}^i\right)$ and six fermionic $\mathcal{G}_{A}^{p}$, usually called charges. The time component $a_t$ of the connection $a\left(t,\varphi\right)$ can be given as:
\begin{eqnarray}\label{bouncond888}
a_t &=&\eta_A\Jt^A +\mu^i \Lt_{i}+\nu_{A}^{p}\Gt_{p}^{A}.
\end{eqnarray}

In this case, we have twelve independent functions $(\eta_A,\mathcal{\mu}^i,\nu_{A}^{p})$, as chemical potentials, which are not allowed to vary, $\delta a_t= 0$.

Using the flatness conditions $(\ref{flatness})$, the equations of motion for fixed chemical potentials impose the following additional conditions on the charges $(\mathcal{J}^A,\mathcal{L}^i, \mathcal{G}_{A}^{p})$:

\begin{eqnarray}
\partial_{t}\mathcal{J}^A	&=& \frac{k}{2}\partial_{\varphi}\eta_A +\epsilon^{ABC}(\mathcal{J}^B\eta_C -\frac{1}{2}(\mathcal{G}_{A}^{\frac{1}{2}}\nu_{C}^{-\frac{1}{2}} + \mathcal{G}_{A}^{-\frac{1}{2}}\nu_{C}^{\frac{1}{2}} )\\
2\partial_{t}\mathcal{L}^{0}&=& -\frac{k}{4}\partial_{\varphi}\mu^{0} -\mathcal{L}^{+1}\mu^{-1} +\mathcal{L}^{-1}\mu^{+1} -\frac{1}{2}(\mathcal{G}_{A}^{\frac{1}{2}}\nu_{A}^{-\frac{1}{2}} -\mathcal{G}_{A}^{-\frac{1}{2}}\nu_{A}^{\frac{1}{2}})\\
\partial_{t}\mathcal{L}^{\pm 1}	&=&	\frac{k}{2}\partial_{\varphi}\mu^{\mp 1} +\mathcal{L}^{0}\mu^{\mp 1} \mp \mathcal{L}^{\mp 1}\mu^{0} \mp\mathcal{G}_{A}^{\mp\frac{1}{2}}\nu_{A}^{\mp \frac{1}{2}},  \\
\partial_{t}\mathcal{G}_{A}^{\mp\frac{1}{2}}&=&	\mp k\partial_{\varphi}\nu_{A}^{\mp\frac{1}{2}} \mp\frac{1}{2} \mu^{0}\mathcal{G}_{A}^{\mp\frac{1}{2}} \mp \mu^{-1}\mathcal{G}_{A}^{\pm\frac{1}{2}} +2\mathcal{L}^{0}\nu_{A}^{\mp\frac{1}{2}} +2\mathcal{L}^{\mp}\nu_{A}^{\pm\frac{1}{2}}\\
&-&\epsilon^{ABC}(\eta_{_{B}} \mathcal{G}_{C}^{\mp\frac{1}{2}} +2\mathcal{J}^B\nu_{C}^{\mp\frac{1}{2}} )
\end{eqnarray}

representing the temporal evolution of the eight state-dependent source fields.\\

We aim to derive the asymptotic symmetry algebra for the most general boundary conditions through canonical analysis. To achieve this, we consider all gauge transformations given by
\begin{equation}\label{boundarycond}
    \delta_{\lambda}\Gamma = \mathrm{d}\lambda + \left[\Gamma, \lambda \right],
\end{equation}
which preserve the most general boundary conditions. At this stage, it is appropriate to express the gauge parameter in terms of the $\mathfrak{osp}(3|2)$ Lie superalgebra basis:
\begin{equation}\label{boundarycond2}
    \lambda = b^{-1}\bigg[\varrho_A\Jt^A + \epsilon^i \Lt_{i} + \zeta_{A}^{p}\Gt_{p}^{A}\bigg]b.
\end{equation}
Note that the gauge parameter includes four bosonic variables $\varrho_A$, $\epsilon^i$, and four fermionic variables $\zeta_{M}^{p}$, which are arbitrary functions of boundary coordinates.

Now, let's focus on the gauge parameters satisfying \eqref{boundarycond}. The infinitesimal gauge transformations are determined as follows:
\begin{align}\label{transf21} 
    \delta_\lambda\mathcal{J}^A	&= \frac{k}{2}\partial_{\varphi}\varrho_A + \epsilon^{ABC}(\mathcal{J}^B\varrho_C - \frac{1}{2}(\mathcal{G}_{A}^{\frac{1}{2}}\zeta_{C}^{-\frac{1}{2}} + \mathcal{G}_{A}^{-\frac{1}{2}}\zeta_{C}^{\frac{1}{2}})),\\
    2\delta_\lambda\mathcal{L}^{0}	&= -\frac{k}{4}\partial_{\varphi}\epsilon^{0} - \mathcal{L}^{+1}\epsilon^{-1} + \mathcal{L}^{-1}\epsilon^{+1} - \frac{1}{2}(\mathcal{G}_{A}^{\frac{1}{2}}\zeta_{A}^{-\frac{1}{2}} - \mathcal{G}_{A}^{-\frac{1}{2}}\zeta_{A}^{\frac{1}{2}}),\\
    \delta_\lambda\mathcal{L}^{\pm 1}	&= \frac{k}{2}\partial_{\varphi}\epsilon^{\mp 1} + \mathcal{L}^{0}\epsilon^{\mp 1} \mp \mathcal{L}^{\mp 1}\epsilon^{0} \mp\mathcal{G}_{A}^{\mp\frac{1}{2}}\zeta_{A}^{\mp \frac{1}{2}},\\
    \delta_\lambda\mathcal{G}_{A}^{\mp\frac{1}{2}}&= \mp k\partial_{\varphi}\zeta_{A}^{\mp\frac{1}{2}} \mp\frac{1}{2} \epsilon^{0}\mathcal{G}_{A}^{\mp\frac{1}{2}} \mp \epsilon^{-1}\mathcal{G}_{A}^{\pm\frac{1}{2}} + 2\mathcal{L}^{0}\zeta_{A}^{\mp\frac{1}{2}} + 2\mathcal{L}^{\mp}\zeta_{A}^{\pm\frac{1}{2}} \nonumber\\
    &- \epsilon^{ABC}(\varrho_{_{B}} \mathcal{G}_{C}^{\mp\frac{1}{2}} + 2\mathcal{J}^B\zeta_{C}^{\mp\frac{1}{2}} )\label{transf22}.
\end{align}
Constraints for the chemical potentials can also be derived analogously:

\begin{align}
    \delta_\lambda\eta^A	&= \frac{k}{2}\partial_{t}\varrho_A + \epsilon^{ABC}(\eta^B\varrho_C - \frac{1}{2}(\mathcal{G}_{A}^{\frac{1}{2}}\zeta_{C}^{-\frac{1}{2}} + \mathcal{G}_{A}^{-\frac{1}{2}}\zeta_{C}^{\frac{1}{2}})),\label{chemicalpotvar1}\\
    2\delta_\lambda\mu^{0}	&= -\frac{k}{4}\partial_{t}\epsilon^{0} - \mu^{+1}\epsilon^{-1} + \mu^{-1}\epsilon^{+1} - \frac{1}{2}(\nu_{A}^{\frac{1}{2}}\zeta_{A}^{-\frac{1}{2}} - \nu_{A}^{-\frac{1}{2}}\zeta_{A}^{\frac{1}{2}}),\\
    \delta_\lambda\mu^{\pm 1}	&= \frac{k}{2}\partial_{t}\epsilon^{\mp 1} + \mu^{0}\epsilon^{\mp 1} \mp \mu^{\mp 1}\epsilon^{0} \mp\nu_{A}^{\mp\frac{1}{2}}\zeta_{A}^{\mp \frac{1}{2}},\\
    \delta_\lambda\nu_{A}^{\mp\frac{1}{2}}&= \mp k\partial_{t}\zeta_{A}^{\mp\frac{1}{2}} \mp\frac{1}{2} \epsilon^{0}\nu_{A}^{\mp\frac{1}{2}} \mp \epsilon^{-1}\nu_{A}^{\pm\frac{1}{2}} + 2\mu^{0}\zeta_{A}^{\mp\frac{1}{2}} + 2\mu^{\mp}\zeta_{A}^{\pm\frac{1}{2}} \nonumber\\
    &- \epsilon^{ABC}(\varrho_{_{B}} \nu_{C}^{\mp\frac{1}{2}} + 2\eta^B\zeta_{C}^{\mp\frac{1}{2}} ).\label{chemicalpotvar2}
 \end{align}

According to the boundary conditions, given in equations \eqref{flatness} - \eqref{conn}, the left-hand side of these equations $(\ref{chemicalpotvar1})$-$(\ref{chemicalpotvar2})$ equals zero. 
 As a final step, the canonical boundary charge $\mathcal{Q[\lambda]}$ that generates the transformations \eqref{transf21} - \eqref{transf22} can be defined. The variation of the canonical boundary charge $\mathcal{Q[\lambda]}$ \cite{Banados:1994tn} leading to the asymptotic symmetry algebra is given by
\begin{equation}\label{Qvar}
    \delta_\lambda \mathcal{Q} = \frac{k}{2\pi}\int\mathrm{d}\varphi\;\mathfrak{str}\left(\lambda \delta \Gamma_{\varphi}\right).
\end{equation}
Therefore, the variation of the canonical boundary charge $\mathcal{Q[\lambda]}$ can be functionally integrated to yield 
\begin{equation}\label{boundaryco9}
    \mathcal{Q[\lambda]} = \int\mathrm{d}\varphi\; \left[\mathcal{J}^A \varrho_A + \mathcal{L}^{i}\epsilon^{-i} + \mathcal{G}_{A}^{p}\zeta_{A}^{-p}\right].
\end{equation}
After determining both the infinitesimal transformations and the canonical boundary charge, we are now positioned to derive the asymptotic symmetry algebra using the standard method \cite{Blgojevic:2002}. This can be obtained through the following relation:
\begin{equation}\label{Qvar2}
    \delta_{\lambda} \digamma = \{\digamma, \mathcal{Q}[\lambda]\}
\end{equation}
for any phase space functional $\digamma$. The charges $\mathcal{L}^i(z)$, $\mathcal{J}^A(z)$, and $\mathcal{G}_{M}^{p}(z)$ are identified as the generators of the asymptotic symmetry algebra. Finally, the operator product algebra can be expressed as follows:
\begin{eqnarray}\label{ope22}
    \mathcal{L}^{i}(z_1)\mathcal{L}^{j}(z_2) & \sim & \frac{\frac{k}{2}\eta^{ij}}{z_{12}^{2}} + \frac{(i-j)}{z_{12}} \mathcal{L}^{i+j}(z_2),\\
    \mathcal{L}^{i}(z_1)\mathcal{G}_{A}^{p}(z_2) & \sim & \frac{({i\over 2}-p)}{z_{12}} \mathcal{G}_{A}^{i+p}(z_2),\\
    \mathcal{J}^A(z_1)\mathcal{J}^B(z_2) & \sim & \frac{\frac{k}{2}\eta}{z_{12}^2}\delta^{AB} + \frac{\epsilon^{ABC}}{z_{12}} \mathcal{J}^{C}(z_2),\\
    \mathcal{J}^A(z_1)\mathcal{G}_{B}^{p}(z_2) & \sim & \frac{\epsilon^{ABC}}{z_{12}} \mathcal{G}_{C}^p(z_2),\\
    \mathcal{G}_{A}^{p}(z_1)\mathcal{G}_{B}^{q}(z_2) & \sim &
    \frac{\frac{k}{2}}{z_{12}^{2}}\eta_{AB}^{pq} + \frac{2}{z_{12}}\delta^{AB} \mathcal{L}^{p+q}(z_2) + \delta_{p,-q}(p-q)\frac{\epsilon^{ABC}}{z_{12}} \mathcal{J}^{C}(z_2).
\end{eqnarray}
Here $\eta^{ij}=\mathfrak{str}(\Lt_i\Lt_j)$, and $\eta^{pq}=\mathfrak{str}(\Gt_p^{A}\Gt_q^{A})$ are the bilinear forms in the fundamental representation of the $\mathfrak{osp}(3|2)$ Lie superalgebra. The operator product algebra can also be expressed in a more compact form:
\begin{eqnarray}\label{ope11}
    \mathfrak{\mathcal{\mathfrak{J}}}^{A}(z_1)\mathfrak{\mathcal{\mathfrak{J}}}^{B}(z_2) & \sim &
    \frac{\frac{k}{2}\eta^{AB}}{z_{12}^{2}} + \frac{\mathfrak{\mathcal{\mathfrak{f}}}^{AB}_{~~~C}
    \mathfrak{\mathcal{\mathfrak{J}}}^{C}(z_2)}{z_{12}}.
\end{eqnarray}
Note that $\eta^{AB}$ is the supertrace matrix, and $\mathfrak{\mathcal{\mathfrak{f}}}^{AB}_{~~C}$ are the structure constants of the related algebra with $(A,B=0,\pm1,\pm\frac{1}{2})$, i.e., $\eta^{ip}=0$ and $\mathfrak{\mathcal{\mathfrak{f}}}^{ij}_{~~i+j}=(i-j)$. Lastly, by repeating the same analysis for the barred sector, the asymptotic symmetry algebra of $\mathcal{N}=(3,3)$ supergravity for the loosest set of boundary conditions is given by two copies of the affine $\mathfrak{osp}(3|2)_k$ algebra.

\subsection{Superconformal Boundary Conditions}\label{bhreduction1}
In this section, our objective is to examine the asymptotic symmetry algebra for the supersymmetric extension of the Brown-Henneaux boundary conditions. We commence by imposing the Drinfeld-Sokolov highest weight gauge condition on the $\mathfrak{osp}(3|2)$ Lie superalgebra-valued connection (\ref{bouncond999}) to further constrain the coefficients. Thus, the Drinfeld-Sokolov reduction sets the fields as follows:
\begin{eqnarray}
\mathcal{L}^0 = \mathcal{G}_{A}^{+\frac{1}{2}} = 0, \quad \mathcal{L}^{-1} = \mathcal{L}, \quad \mathcal{G}_{A}^{-\frac{1}{2}} = \mathcal{G}_{A}, \quad \gamma_{+1}\mathcal{L}^{+1} = 1.
\end{eqnarray}
It is noteworthy that the superconformal boundary conditions are the supersymmetric extension of the well-known Brown-Henneaux boundary conditions proposed in \cite{Brown:1986nw} for $AdS_3$ supergravity. Consequently, the supersymmetric gauge connection takes the form:
\begin{eqnarray}\label{bouncondhf}
a_\varphi &=& \Lt_{+1} + \gamma_{-1}\mathcal{L} \Lt_{-1} + \rho_A\mathcal{J}^A\Jt^A + \sigma_{A}^{-{\frac{1}{2}}}\mathcal{G}_{A}\Gt_{-{\frac{1}{2}}}^{A}.
\end{eqnarray}
Here, $\gamma_{-1} = -\frac{1}{k}$, $\rho = \frac{1}{2k}$, $\sigma_{+}^{-{\frac{1}{2}}} = \frac{1}{2k}$, and $\sigma_{-}^{-{\frac{1}{2}}} = -\frac{1}{2k}$ are scaling parameters. We introduce four functions: two bosonic ($\mathcal{J},\mathcal{L}$) and two fermionic ($\mathcal{G}_{\pm}$) charges. After implementing these steps, we are on the verge of obtaining the superconformal asymptotic symmetry algebra. Following the results implied by the Drinfeld-Sokolov reduction, the gauge parameter $\lambda$ has only four independent functions $(\varrho_A,\mathcal{\epsilon}\equiv\epsilon^{+1},\,\zeta_{A}\equiv \zeta_{A}^{+\frac{1}{2}})$, given by:
\begin{eqnarray}\label{ebc08}
\lambda &=& b^{-1}\left[
  \epsilon \Lt_{1}
 -\epsilon'\Lt_{0}
+ \Bigg( \frac{6\mathcal{L}\epsilon}{c}
       +\frac{3 \mathcal{G}_A \zeta_A}{c}
       +\frac{\epsilon ''}{2}
       \Bigg)\Lt_{-1}
       \right.\nonumber\\
&&\left.
+ \varrho_A \Jt^A
+ \zeta_A \Gt_{\frac{1}{2}}^A
+ \left(\frac{3 \epsilon^{ABC}\mathcal{J}^B \zeta_C}{c}-\frac{3 \mathcal{G}_A \epsilon }{c}-\zeta_A'\right) \Gt_{-\frac{1}{2}}^A
\right]b.
\end{eqnarray}
Substituting this gauge parameter into the expression for the transformation of the fields (\ref{boundarycond}), we obtain the infinitesimal gauge transformations:
\begin{eqnarray}\label{ebc086}
\delta_{\lambda}\mathcal{L}&=&
\frac{c}{12}  \epsilon ^{'''}
+\epsilon  \mathcal{L}'+2 \mathcal{L} \epsilon'
+\frac{3}{2} \mathcal{G}_A \zeta_A'
+\frac{1}{2} \zeta_A \mathcal{G}_A'
+\frac{3 \epsilon^{ABC}\zeta_A\mathcal{J}^B\mathcal{G}_C }{c}\\
\delta_{\lambda}\mathcal{J}^A&=&
-\frac{c}{3}\varrho _A'
+\epsilon^{ABC}\bigg(\mathcal{J}^B\varrho _C+\mathcal{G}_B \zeta_C\bigg)\\
\delta_{\lambda}\mathcal{G}^A&=&
\frac{c}{3}\zeta_A''+2\bigg(\mathcal{L}-\frac{3}{2c}\mathcal{J}^B\mathcal{J}^B\bigg)\zeta_A
+ \frac{3\mathcal{J}^A\mathcal{J}^B\zeta_B}{c}
+\frac{3\mathcal{G}_A\epsilon'}{2}
+\epsilon\mathcal{G}_A'\nonumber\\
&+& \epsilon^{ABC}\Bigg(\bigg(\varrho_C+\frac{3}{c}\epsilon\mathcal{J}^C\bigg)\mathcal{G}_B+\zeta_B{\mathcal{J}^C}'-2\mathcal{J}^B\zeta_C'\Bigg)
\end{eqnarray}

For the given boundary conditions, the boundary charge $\mathcal{Q}[\lambda]$ is integrable, and we identify $\mathcal{L}$, $\mathcal{G}^A$, and $\mathcal{J}^A$ as charges. However, these charges do not generate an algebra due to some nonlinear terms $(\mathcal{J}^B \mathcal{G}_C)$ in $(\ref{ebc086})$. Moreover, to obtain the correct superconformal algebra $(\ref{qalg11})$-$(\ref{qalg12})$, we need to perform a shift on $\mathcal{L}$ and a redefinition of the gauge parameter $\varrho_A$ as:
\begin{equation}\label{shift1}
  \mathcal{L} \rightarrow \mathcal{L} + \frac{3}{2c} \mathcal{J}^A \mathcal{J}^A, \quad \varrho_A \rightarrow \varrho_A - \frac{3}{c}\epsilon \mathcal{J}^A.
\end{equation}
The new gauge parameter $\lambda$ is given as:
\begin{eqnarray}\label{ebc08}
  \lambda &=& b^{-1}\left[\epsilon \Lt_{1} - \epsilon'\Lt_{0} + \Bigg( \frac{6}{c} \bigg(\mathcal{L}+\frac{3}{2c} \mathcal{J}^A\mathcal{J}^A\bigg) \epsilon + \frac{3 \mathcal{G}_A \zeta_A}{c} + \frac{\epsilon ''}{2} \Bigg)\Lt_{-1}\right.\nonumber\\
  &&\left.+ \left(\varrho_A - \frac{3 \mathcal{J}^A \epsilon }{c}\right) \Jt^A + \zeta_A \Gt_{\frac{1}{2}}^A + \left(-\frac{3 \mathcal{G}_A \epsilon }{c} + \frac{3 \epsilon^{ABC}\mathcal{J}^B \zeta_C}{c} - \zeta_A'\right) \Gt_{-\frac{1}{2}}^A\right]b.
\end{eqnarray}
Substituting this new gauge parameter into the expression $(\ref{boundarycond})$, we obtain the new infinitesimal gauge transformations:
\begin{eqnarray}\label{ebc08}
  \delta_{\lambda}\mathcal{L} &=& \frac{c}{12}  \epsilon ^{'''} + \epsilon  \mathcal{L}' + 2 \mathcal{L} \epsilon' + \frac{3}{2} \mathcal{G}_A \zeta_A' + \frac{1}{2} \zeta_A \mathcal{G}_A' + \frac{3 \epsilon^{ABC}\zeta_A\mathcal{J}^B\mathcal{G}_C }{c},\\
  \delta_{\lambda}\mathcal{J}^A &=& -\frac{c}{3}\varrho _A' + \epsilon^{ABC}\bigg(\mathcal{J}^B\varrho _C+\mathcal{G}_B \zeta_C\bigg),\\
  \delta_{\lambda}\mathcal{G}^A &=& \frac{c}{3}\zeta_A''+2\mathcal{L}\zeta_A + \frac{3\mathcal{J}^A\mathcal{J}^B\zeta_B}{c} + \frac{3\mathcal{G}_A\epsilon'}{2} + \epsilon\mathcal{G}_A'\nonumber\\
  &&+ \epsilon^{ABC}\Bigg(\varrho_C\mathcal{G}_B+\zeta_B{\mathcal{J}^C}'-2\mathcal{J}^B\zeta_C'\Bigg).
\end{eqnarray}
In fact, these gauge transformations provide insight into the asymptotic symmetry algebra \cite{Banados:1998pi}. By examining the asymptotic symmetries, one can integrate the variation of the canonical boundary charges, i.e., $\delta_\lambda \mathcal{Q}$ in expression $(\ref{Qvar})$, such that
\begin{equation}
  \mathcal{Q}[\lambda] = \int\mathrm{d}\varphi\;\left[\mathcal{L}\epsilon + \mathcal{G}_A\zeta_A + \mathcal{J}^A\varrho_A\right].\label{boundaryco11111333333}
\end{equation}
These canonical boundary charges provide a convenient asymptotic operator product algebra for $\mathcal{N}=(3,3)$ superconformal boundaries:
\begin{eqnarray}\label{ope2}
  \mathcal{L}(z_1)\mathcal{L}(z_2) &\sim& \frac{{c/2}}{{z_{12}^{4}}} + \frac{2\mathcal{L}}{{z_{12}^{2}}} + \frac{\mathcal{L}'}{z_{12}},\\
  \mathcal{L}(z_1)\mathcal{J}(z_2) &\sim& \frac{\mathcal{J}}{z_{12}^{2}} + \frac{\mathcal{J}'}{z_{12}}, \\
  \mathcal{L}(z_1)\mathcal{G}_{A}(z_2) &\sim& \frac{{\frac{3}{2}\mathcal{G}_{A}}}{z_{12}^{2}} + \frac{{\mathcal{G}_{A}}'}{z_{12}},\\
  \mathcal{J}^A(z_1)\mathcal{J}^B(z_2) &\sim& \frac{\frac{c}{3}}{z_{12}^2}\delta^{AB} + \frac{\epsilon^{ABC}}{z_{12}}\mathcal{J}^{C} ,\\
  \mathcal{J}^A(z_1)\mathcal{G}_{A}(z_2) &\sim& \frac{\epsilon^{ABC}}{z_{12}}\mathcal{G}_{C},\\
  \mathcal{G}_{A}(z_1)\mathcal{G}_{B}(z_2) &\sim& \delta^{AB}\bigg(\frac{{2c/3}}{{z_{12}^{3}}} + \frac{2\mathcal{L}}{z_{12}}\bigg) \nonumber\\
  &&+ 2\epsilon^{ABC}\bigg(\frac{\mathcal{J}^C}{z_{12}^{2}} + \frac{{\frac{1}{2}\partial\mathcal{J}^C}}{z_{12}}\bigg) + \frac{{\frac{3}{c}\mathcal{J}^A}\mathcal{J}^B}{z_{12}}.
\end{eqnarray}
Upon repeating the same analysis for the barred sector, it is observed that the asymptotic symmetry algebra for the loosest set of boundary conditions of $\mathcal{N}=(3,3)$ supergravity consists of two copies of the super-Virasoro algebra with central charge $c=3k$.

\section{$\mathcal{N}=(4,4)$ $\mathfrak{osp}(4|2) \oplus \mathfrak{osp}(4|2)$ Higher-Spin Chern-Simons Supergravity}

After laying the groundwork with the canonical analysis of the $\mathfrak{osp}(3|2)$ case, we present the extended $\mathcal{N}=(4,4)$ higher-spin Chern-Simons supergravity theory based on the $\mathfrak{osp}(4|2)_k$ superalgebra.
\subsection{Affine Boundary Conditions}

The objective of this section is to construct $\mathcal{N}=(4,4)$ extended higher-spin $AdS_3$ supergravity as a $\mathfrak{osp}(4|2) \oplus \mathfrak{osp}(4|2)$ Chern-Simons gauge theory on the affine boundary. We proceed with our calculations to elucidate the asymptotic symmetry algebra for the loosest set of boundary conditions. As stated in the previous section, we consider the principal embedding of $\mathfrak{osp}(1|2)$ into $\mathfrak{osp}(4|2)$ as a sub-algebra.

We can now formulate the most general boundary conditions for asymptotically $AdS_3$ spacetimes. To achieve this, it is useful to define the gauge connection as follows,

\begin{align}\label{bouncondaf}
    a_\varphi &= \rho_A\mathcal{J}^A\Jt_A
    +\gamma_i\mathcal{L}^i \Lt_{i}
    +\sigma_A^p\mathcal{G}_{A}^{p}\Gt_{p}^{A}
    +\bar{\sigma}_A^p\bar{\mathcal{G}}_{A}^{p}\bar{\Gt}_{p}^{A} \\
    a_t &= \eta_A\Jt_A
    +\mu^i \Lt_{i}
    +\nu_A^p\Gt_{p}^{A}
    +\bar{\nu}_A^p\bar{\Gt}_{p}^{A}
\end{align}

where

\begin{align}
    \rho &= \sigma_M^{\frac{1}{2}} = -\sigma_M^{\frac{1}{2}} = \frac{1}{k}, \quad 2\gamma_1 = 2\gamma_{-1}-2\gamma_0 = \frac{4}{k} 
\end{align}

are scaling parameters. Consequently, we have fourteen functions: six bosonic $\left(\mathcal{J}^A,\mathcal{L}^i\right)$ and eight $\left(\mathcal{G}_{A}^{p}, \bar{\mathcal{G}}_{A}^{p}\right)$ as charges. Additionally, we have a total of fourteen independent functions $\left( \eta_A,\mu^i,\nu_A^p ,\bar{\nu}_A^p\right)$ as chemical potentials for the time component. In the presence of the loosest set of boundary conditions, thanks to the flatness conditions $(\ref{flatness})$, the equations of motion for fixed chemical potentials impose additional conditions as the temporal evolution of the fourteen independent source fields $\left( \mathcal{J}^A ,\mathcal{L}^i,\mathcal{G}_{A}^{p} ,\bar{\mathcal{G}}_{A}^{p} \right)$ as calculated below:

\begin{align}
    \partial_{t}\mathcal{J}^A	&=
    \frac{k}{2}\partial_{\varphi}\eta_A
    +\epsilon^{ABC}\mathcal{J}^B\eta_C \nonumber\\
    &-(\sigma^{A})_{BC}
    \bigg(
    \nu_{B}^{\frac{1}{2}}\bar{\mathcal{G}}_{C}^{-\frac{1}{2}}
    +\nu_{B}^{-\frac{1}{2}}\bar{\mathcal{G}}_{C}^{\frac{1}{2}}
    -\bar{\nu}_{B}^{\frac{1}{2}} {\mathcal{G}}_{C}^{-\frac{1}{2}}
    -\bar{\nu}_{B}^{-\frac{1}{2}}\mathcal{G}_{C}^{\frac{1}{2}}
    \bigg) \\
    2\partial_{t}\mathcal{L}^{0} &=
    -\frac{k}{4}\partial_{\varphi}\mu^{0}
    -\mathcal{L}^{+1}\mu^{-1}
    +\mathcal{L}^{-1}\mu^{+1} \nonumber \\
    &+\frac{1}{2}
    \bigg(
    \bar{\mathcal{G}}_{A}^{\frac{1}{2}}\nu_{A}^{-\frac{1}{2}}
    -\bar{\mathcal{G}}_{A}^{-\frac{1}{2}}\nu_{A}^{\frac{1}{2}}
    - \mathcal{G}_{A}^{-\frac{1}{2}}\nu_{A}^{\frac{1}{2}}
    +\mathcal{G}_{A}^{\frac{1}{2}}\nu_{A}^{-\frac{1}{2}}
    \bigg) \\
    \partial_{t}\mathcal{L}^{\mp 1}	&=	
    \frac{k}{2}\partial_{\varphi}\mu^{\pm 1}
    +2\mathcal{L}^{0}\mu^{\pm 1}
    \pm \mathcal{L}^{\pm 1}\mu^{0}
    \pm\bar{\mathcal{G}}_{A}^{\pm\frac{1}{2}}\nu_{A}^{\pm \frac{1}{2}}
    \pm\mathcal{G}_{A}^{\pm\frac{1}{2}}\bar{\nu}_{A}^{\pm \frac{1}{2}},  \\
    \partial_{t}\mathcal{G}_{A}^{\pm\frac{1}{2}}&=	
    \pm k\partial_{\varphi}\nu_{A}^{\pm\frac{1}{2}}
    \pm\frac{1}{2} \mu^{0}\mathcal{G}_{A}^{\pm\frac{1}{2}}
    \pm\mu^{\mp}\mathcal{G}_{A}^{\pm\frac{1}{2}}
    +2\mathcal{L}^{0}\nu_{A}^{\pm\frac{1}{2}}
    +2\mathcal{L}^{\mp}\nu_{A}^{\pm\frac{1}{2}} \nonumber \\
    &+\frac{i}{2}(\sigma^B)^{*}_{AC}\bigg(\eta_{B} \mathcal{G}_{C}^{\pm\frac{1}{2}}
    +2\mathcal{J}^B\nu_{C}^{\pm\frac{1}{2}} \bigg) \\
    \partial_{t}\bar{\mathcal{G}}_{A}^{\pm\frac{1}{2}}&=	
    \pm k\partial_{\varphi}\nu_{A}^{\pm\frac{1}{2}}
    \pm\frac{1}{2} \mu^{0}\bar{\mathcal{G}}_{A}^{\pm\frac{1}{2}}
    \pm\mu^{\mp}\bar{\mathcal{G}}_{A}^{\pm\frac{1}{2}}
    +2\mathcal{L}^{0}\bar{\nu}_{A}^{\pm\frac{1}{2}}
    +2\mathcal{L}^{\mp}\bar{\nu}_{A}^{\pm\frac{1}{2}} \nonumber \\
    &+\frac{i}{2}(\sigma^B)^{*}_{AC}\bigg(\eta_{B} \bar{\mathcal{G}}_{C}^{\pm\frac{1}{2}}
    -2\mathcal{J}^B\bar{\nu}_{C}^{\pm\frac{1}{2}} \bigg)
\end{align}

Thus, the consequences of our calculations to derive the relevant superalgebra for the loosest set of boundary conditions can now be evaluated through a canonical analysis. We now consider the boundary-preserving gauge transformations (encompassing all) \eqref{boundarycond} generated by the $\mathfrak{osp}(4|2)$ Lie superalgebra-valued gauge parameter $\lambda$, which we choose as
\begin{equation}
    \lambda = b^{-1}
                     \bigg[
                     \varrho_A\Jt_A
                     +\epsilon^i \Lt_{i}
                     +\zeta_A^p\Gt_{p}^{A}
                     +\bar{\zeta}_A^p\bar{\Gt}_{p}^{A}
                     \bigg]b.\label{boundarycond2221}
\end{equation}

Note that there are in total fourteen arbitrary functions on the boundary, consisting of six bosonic $(\varrho_A, \, \epsilon^i)$ and eight fermionic $(\mathcal{\zeta}_{A}^{p}, \, \bar{\mathcal{\zeta}}_{A}^{p})$. Inserting this expression into \eqref{boundarycond} imposes the gauge transformations as:
\begin{align}
    \delta_{\lambda}\mathcal{J}^A & = \frac{k}{2}\partial_{\varphi}\varrho_A + \epsilon^{ABC}\mathcal{J}^B\varrho_C \nonumber \\
    & \quad - (\sigma^{A})_{BC} \bigg( \zeta_{B}^{\frac{1}{2}}\bar{\mathcal{G}}_{C}^{-\frac{1}{2}} + \zeta_{B}^{-\frac{1}{2}}\bar{\mathcal{G}}_{C}^{\frac{1}{2}} \nonumber \\
    & \quad - \bar{\zeta}_{B}^{\frac{1}{2}} {\mathcal{G}}_{C}^{-\frac{1}{2}} - \bar{\zeta}_{B}^{-\frac{1}{2}}\mathcal{G}_{C}^{\frac{1}{2}} \bigg),\label{trans231}\\
    2\delta_{\lambda}\mathcal{L}^{0} & = -\frac{k}{4}\partial_{\varphi}\epsilon^{0} - \mathcal{L}^{+1}\epsilon^{-1} + \mathcal{L}^{-1}\epsilon^{+1} \nonumber \\
    & \quad + \frac{1}{2} \bigg( \bar{\mathcal{G}}_{A}^{\frac{1}{2}}\zeta_{A}^{-\frac{1}{2}} - \bar{\mathcal{G}}_{A}^{-\frac{1}{2}}\zeta_{A}^{\frac{1}{2}} \nonumber \\
    & \quad - \mathcal{G}_{A}^{-\frac{1}{2}}\zeta_{A}^{\frac{1}{2}} + \mathcal{G}_{A}^{\frac{1}{2}}\zeta_{A}^{-\frac{1}{2}} \bigg),\\
    \delta_{\lambda}\mathcal{L}^{\mp 1} & = \frac{k}{2}\partial_{\varphi}\epsilon^{\pm 1} + 2\mathcal{L}^{0}\epsilon^{\pm 1} \nonumber \\
    & \quad \pm \mathcal{L}^{\pm 1}\epsilon^{0} \pm \bar{\mathcal{G}}_{A}^{\pm\frac{1}{2}}\zeta_{A}^{\pm \frac{1}{2}} \nonumber \\
    & \quad \pm \mathcal{G}_{A}^{\pm\frac{1}{2}}\bar{\zeta}_{A}^{\pm \frac{1}{2}},  \\
    \delta_{\lambda}\mathcal{G}_{A}^{\pm\frac{1}{2}} & = \pm k\partial_{\varphi}\zeta_{A}^{\pm\frac{1}{2}} \pm \frac{1}{2} \epsilon^{0}\mathcal{G}_{A}^{\pm\frac{1}{2}} \nonumber \\
    & \quad \pm \epsilon^{\mp}\mathcal{G}_{A}^{\pm\frac{1}{2}} + 2\mathcal{L}^{0}\zeta_{A}^{\pm\frac{1}{2}} + 2\mathcal{L}^{\mp}\zeta_{A}^{\pm\frac{1}{2}} \nonumber \\
    & \quad + \frac{i}{2}(\sigma^B)^{*}_{AC}\bigg(\varrho_{B} \mathcal{G}_{C}^{\pm\frac{1}{2}} + 2\mathcal{J}^B\zeta_{C}^{\pm\frac{1}{2}} \bigg),\\
    \delta_{\lambda}\bar{\mathcal{G}}_{A}^{\pm\frac{1}{2}} & = \pm k\partial_{\varphi}\zeta_{A}^{\pm\frac{1}{2}} \pm \frac{1}{2} \epsilon^{0}\bar{\mathcal{G}}_{A}^{\pm\frac{1}{2}} \nonumber \\
    & \quad \pm \epsilon^{\mp}\bar{\mathcal{G}}_{A}^{\pm\frac{1}{2}} + 2\mathcal{L}^{0}\bar{\zeta}_{A}^{\pm\frac{1}{2}} \nonumber \\
    & \quad + 2\mathcal{L}^{\mp}\bar{\zeta}_{A}^{\pm\frac{1}{2}} + \frac{i}{2}(\sigma^B)^{*}_{AC}\bigg(\varrho_{B} \bar{\mathcal{G}}_{C}^{\pm\frac{1}{2}} \nonumber \\
    & \quad - 2\mathcal{J}^B\bar{\zeta}_{C}^{\pm\frac{1}{2}} \bigg)\label{trans232}
\end{align}

Analogously, the gauge transformations for the chemical potentials are calculated as:
\begin{align}
    \delta_{\lambda}\eta^A & = \frac{k}{2}\partial_{t}\varrho_A + \epsilon^{ABC}\eta^B\varrho_C \nonumber \\
    & \quad - (\sigma^{A})_{BC} \bigg( \zeta_{B}^{\frac{1}{2}}\bar{\nu}_{C}^{-\frac{1}{2}} + \zeta_{B}^{-\frac{1}{2}}\bar{\nu}_{C}^{\frac{1}{2}} \nonumber \\
    & \quad - \bar{\zeta}_{B}^{\frac{1}{2}} {\nu}_{C}^{-\frac{1}{2}} - \bar{\zeta}_{B}^{-\frac{1}{2}}\nu_{C}^{\frac{1}{2}} \bigg),\\
    2\delta_{\lambda}\mu^{0} & = -\frac{k}{4}\partial_{t}\epsilon^{0} - \mu^{+1}\epsilon^{-1} + \mu^{-1}\epsilon^{+1} \nonumber \\
    & \quad + \frac{1}{2} \bigg( \bar{\nu}_{A}^{\frac{1}{2}}\zeta_{A}^{-\frac{1}{2}} - \bar{\nu}_{A}^{-\frac{1}{2}}\zeta_{A}^{\frac{1}{2}} \nonumber \\
    & \quad - \nu_{A}^{-\frac{1}{2}}\zeta_{A}^{\frac{1}{2}} + \nu_{A}^{\frac{1}{2}}\zeta_{A}^{-\frac{1}{2}} \bigg),\\
    \delta_{\lambda}\mu^{\mp 1} & = \frac{k}{2}\partial_{t}\epsilon^{\pm 1} + 2\mu^{0}\epsilon^{\pm 1} \nonumber \\
    & \quad \pm \mu^{\pm 1}\epsilon^{0} \pm \bar{\nu}_{A}^{\pm\frac{1}{2}}\zeta_{A}^{\pm \frac{1}{2}} \nonumber \\
    & \quad \pm \nu_{A}^{\pm\frac{1}{2}}\bar{\zeta}_{A}^{\pm \frac{1}{2}},  \\
    \delta_{\lambda}\nu_{A}^{\pm\frac{1}{2}} & = \pm k\partial_{t}\zeta_{A}^{\pm\frac{1}{2}} \pm \frac{1}{2} \epsilon^{0}\nu_{A}^{\pm\frac{1}{2}} \nonumber \\
    & \quad \pm \epsilon^{\mp}\nu_{A}^{\pm\frac{1}{2}} + 2\mu^{0}\zeta_{A}^{\pm\frac{1}{2}} + 2\mu^{\mp}\zeta_{A}^{\pm\frac{1}{2}} \nonumber \\
    & \quad + \frac{i}{2}(\sigma^B)^{*}_{AC}\bigg(\varrho_{B} \nu_{C}^{\pm\frac{1}{2}} + 2\eta^B\zeta_{C}^{\pm\frac{1}{2}} \bigg),\\
    \delta_{\lambda}\bar{\nu}_{A}^{\pm\frac{1}{2}} & = \pm k\partial_{t}\zeta_{A}^{\pm\frac{1}{2}} \pm \frac{1}{2} \epsilon^{0}\bar{\nu}_{A}^{\pm\frac{1}{2}} \nonumber \\
    & \quad \pm \epsilon^{\mp}\bar{\nu}_{A}^{\pm\frac{1}{2}} + 2\mu^{0}\bar{\zeta}_{A}^{\pm\frac{1}{2}} \nonumber \\
    & \quad + 2\mu^{\mp}\bar{\zeta}_{A}^{\pm\frac{1}{2}} + \frac{i}{2}(\sigma^B)^{*}_{AC}\bigg(\varrho_{B} \bar{\nu}_{C}^{\pm\frac{1}{2}} \nonumber \\
    & \quad - 2\eta^B\bar{\zeta}_{C}^{\pm\frac{1}{2}} \bigg)
\end{align}

Following a methodology analogous to the previous section, we proceed to identify the canonical boundary charges $\mathcal{Q}[\lambda]$ responsible for generating the transformations $(\ref{trans231})$-$(\ref{trans232})$. As is well-known, it is convenient to express the variation of the canonical boundary charge $\delta_\lambda \mathcal{Q}$ (see $(\ref{Qvar})$) to deduce the asymptotic symmetry algebra \cite{Banados:1994tn}. Therefore, the canonical boundary charge $\mathcal{Q}[\lambda]$ is derived as follows:
\begin{equation}\label{boundaryco13}
\mathcal{Q}[\lambda] = \int\mathrm{d}\varphi\;\left[
 \mathcal{J}^A \varrho_A
+\mathcal{L}^{i}\epsilon^{-i}
+\mathcal{G}_{A}^{p}\zeta_{A}^{-p}
+\bar{\mathcal{G}}_{A}^{p}\bar{\zeta_{A}}^{-p}
\right].
\end{equation}

The subsequent step towards establishing the asymptotic symmetry algebra involves calculating the Poisson bracket algebra using the standard method \cite{Blgojevic:2002}, as obtained from the relation $(\ref{Qvar2})$, which holds for any phase space functional $\digamma$.

In light of the aforementioned, the operator product algebra for the bosonic sector is then expressed as follows:
\begin{align}
\label{ope22}
\mathcal{L}^{i}(z_1)\mathcal{L}^{j}(z_2)\,& \sim \, \frac{\frac{k}{2}\eta^{ij}}{z_{12}^{2}}\,+\,\frac{(i-j)}{z_{12}} \mathcal{L}^{i+j} ,\\
\mathcal{L}^{i}(z_1)\mathcal{G}_{A}^{p}(z_2)\,& \sim \,\frac{({i\over 2}-p)}{z_{12}} \mathcal{G}_{A}^{i+p} ,\\
\mathcal{L}^{i}(z_1)\bar{\mathcal{G}}_{A}^{p}(z_2)\,& \sim \,\frac{({i\over 2}-p)}{z_{12}}\bar{ \mathcal{G}}_{A}^{i+p} ,\\
\mathcal{J}^A(z_1)\mathcal{J}^B(z_2)\,& \sim \, \frac{\frac{k}{2}\delta^{AB}}{z_{12}^2}+\,\frac{\epsilon^{ABC}}{z_{12}} \mathcal{J}^{C} ,\\
\mathcal{J}^A(z_1)\mathcal{G}_{B}^{p}(z_2)\,& \sim \,\frac{(\sigma^A)_{BC}}{z_{12}} \bar{\mathcal{G}}_{C}^p ,\\
\mathcal{J}^A(z_1)\bar{\mathcal{G}}_{B}^{p}(z_2)\,& \sim \,\frac{(\sigma^A)_{BC}}{z_{12}}  \mathcal{G}_{C}^p ,\\
\mathcal{G}_{A}^{p}(z_1)\mathcal{G}_{B}^{q}(z_2)\,& \sim \,
\delta^{AB}\left(\frac{\frac{k}{2}}{z_{12}^{2}}\eta_{AB}^{pq}\,+\,\frac{2}{z_{12}} \mathcal{L}^{p+q}\right) \nonumber \\
&-i \delta_{p,-q}(p-q)\frac{(\sigma^{C})_{AB}}{z_{12}} \mathcal{J}^{C}.
\end{align}
or in a more concise form,
\begin{equation}\label{ope111}
\mathfrak{\mathcal{\mathfrak{J}}}^{A}(z_1)\mathfrak{\mathcal{\mathfrak{J}}}^{B}(z_2)\,\sim\, \frac{\frac{k}{2}\eta^{AB}}{z_{12}^{2}}\,+\,\frac{\mathfrak{\mathcal{\mathfrak{f}}}^{AB}_{~~~C} \mathfrak{\mathcal{\mathfrak{J}}}^{C}(z_2)}{z_{12}}.
\end{equation}
Note that $\eta^{AB}$ is the supertrace matrix, and $\mathfrak{\mathcal{\mathfrak{f}}}^{AB}_{~~C}$'s are the structure constants of the related algebra with $(A,B=0,\pm1,\pm\frac{1}{2},0,\pm1,\pm\frac{1}{2},\pm\frac{3}{2})$, i.e., $\eta^{ip}=0$ and $\mathfrak{\mathcal{\mathfrak{f}}}^{ij}_{~~i+j}=(i-j)$.

Since the barred sector is entirely analogous, identical results are obtained. Consequently, it is deduced that the asymptotic symmetry algebra for the loosest set of boundary conditions of $\mathcal{N}=(4,4)$ supergravity is two copies of the affine $\mathfrak{osp}(4|2)_k$ algebra.

\subsection{Superconformal Boundary Conditions}

As discussed previously, it is convenient to note that superconformal boundary conditions are the supersymmetric extension of the well-known Brown-Henneaux boundary conditions presented in \cite{Brown:1986nw} for $AdS_3$ supergravity. In this section, our primary objective is to construct the asymptotic symmetry algebra for the most general boundary conditions, serving as the supersymmetric extension of the Brown-Henneaux conditions. To achieve this, we initiate our discussion by imposing the Drinfeld-Sokolov highest weight gauge condition on the $\mathfrak{osp}(3|2)$ Lie superalgebra-valued connection (\ref{bouncondaf}), setting the fields as follows:

\begin{eqnarray}\label{bouncondaysinss}
    \mathcal{L}^0 &=& \mathcal{G}_{M}^{+\frac{1}{2}} = \bar{\mathcal{G}}_{M}^{+\frac{1}{2}} = 0,\nonumber\\
    \mathcal{L}^{-1} &=& \mathcal{L}, \quad \mathcal{G}_{M}^{-\frac{1}{2}} = \mathcal{G}_{M}, \quad \bar{\mathcal{G}}_{M}^{-\frac{1}{2}} = \bar{\mathcal{G}}_{M}, \quad \gamma_{+1}\mathcal{L}^{+1} = 1.
\end{eqnarray}

Correspondingly, the supersymmetric gauge connection is given by:

\begin{eqnarray}\label{bouncondaysin}
    a_\varphi &=& \Lt_{1} + \gamma_{-1}\mathcal{L} \Lt_{-1} + \rho_A\mathcal{J}^A\Jt_A \nonumber\\
    & & + \sigma_M^{-{\frac{1}{2}}}\mathcal{G}_{M}\Gt_{-{\frac{1}{2}}}^{M} + \bar{\sigma}_M^{-{\frac{1}{2}}}\bar{\mathcal{G}}_{M}\bar{\Gt}_{-{\frac{1}{2}}}^{M},\\
    a_t &=& \eta_A\Jt_A + \mu \Lt_{1} + \nu_{M}\Gt_{{+\frac{1}{2}}}^{M} + \bar{\nu}_{M}\bar{\Gt}_{{+\frac{1}{2}}}^{M}\nonumber\\
    & & + \sum_{i=-1}^{0}\mu^i \Lt_{i} + \nu_{M}^{-{\frac{1}{2}}}\Gt_{-{\frac{1}{2}}}^{M} + \bar{\nu}_{M}^{-{\frac{1}{2}}}\bar{\Gt}_{-{\frac{1}{2}}}^{M},
\end{eqnarray}

where $\eta_A$, $\mu\equiv \mu^{+1}$, $\nu_{_M}\equiv \nu_{_M}^{+\frac{1}{2}}$, and $\bar{\nu}_{_M}\equiv \bar{\nu}_{_M}^{+\frac{1}{2}}$ are interpreted as the independent chemical potentials.

When summarizing our next steps towards deriving the asymptotic symmetry algebra in one paragraph, it is appropriate to provide the following brief overview. All functions except the chemical potentials can be determined by the flatness conditions (\ref{flatness}) in the usual manner. The equations of motion for the fixed chemical potentials can also be obtained conventionally as the time evolution of the canonical boundary charges. However, the extension to $\mathfrak{osp}(4|2)$ introduces technical complexities, which, though overcome, are omitted here due to space constraints. We choose to allocate space for the calculations of the gauge parameter $\lambda$ and further developments.

Based on the results to be obtained, we can now derive the superconformal asymptotic symmetry algebra. Using the Drinfeld-Sokolov reduction, we have only six independent parameters: $\mathcal{\varrho_A}$, $\mathcal{\epsilon}\equiv\mathcal{\epsilon}^{+1}$, $\mathcal{\zeta}_{_M}\equiv\mathcal{\zeta}_{_M}^{+{\frac{1}{2}}}$, and $\bar{\mathcal{\zeta}}_{_M}\equiv\bar{\mathcal{\zeta}}_{_M}^{+{\frac{1}{2}}}$. It is now possible to compute the gauge transformations by considering all transformations (\ref{boundarycond}) that preserve the boundary conditions with the $\mathfrak{osp}(4|2)$ Lie superalgebra-valued gauge parameter $\lambda$.\\

\begin{eqnarray}\label{ebc08}
\lambda &=& b^{-1}\left[
  \epsilon \Lt_{1}
 -\epsilon'\Lt_{0}
+ \left( \frac{6}{c} \mathcal{L} \epsilon
       +\frac{3 \bar{\mathcal{G}}_A \zeta_A}{c}
       +\frac{3 \mathcal{G}_A \bar{\zeta_A}}{c}
       +\frac{\epsilon ''}{2}
       \right)\Lt_{-1}
       \right.\nonumber\\
&&\left.
+  \varrho_A  \Jt^A
+  \zeta_A  \Gt_{\frac{1}{2}}^A
+  \bar{\zeta}_A  \bar{\Gt}_{\frac{1}{2}}^A
+\left(\frac{6 (\sigma^A)^{*}_{BC}\mathcal{J}^A \zeta_C}{c}-\frac{3 \mathcal{G}_A \epsilon }{c}-\zeta_A'\right) \Gt_{-\frac{1}{2}}^A
\right.\nonumber\\
&&\left.
-\left(\frac{6 (\sigma^A)_{BC}\mathcal{J}^A \bar{\zeta_C}}{c}+\frac{3 \mathcal{G}_A \epsilon }{c}+\bar{\zeta_A}'\right) \bar{\Gt}_{-\frac{1}{2}}^A
\right]b.
\end{eqnarray}

Since we are dealing with asymptotic symmetries, it is natural to demand obtaining the canonical boundary charges $\mathcal{Q}[\lambda]$. So, the variation of the canonical boundary charge, i.e., $\delta_\lambda \mathcal{Q}$ (Eq. (\ref{Qvar})), can be integrated to yield
\begin{equation}
\mathcal{Q}[\lambda]=\int\mathrm{d}\varphi\;
\left[
 \mathcal{J}^A\varrho_A
+\mathcal{L}\epsilon
+\mathcal{G}_{_M}\zeta_{_M}
+\bar{\mathcal{G}}_{_M}\bar{\zeta}_{_M}
\right].
\label{boundaryco11111}
\end{equation}

However, these canonical boundary charges do not yield a convenient asymptotic operator product algebra in the complex coordinates when using Eq. (\ref{Qvar2}) for $\mathcal{N}=(4,4)$ superconformal boundaries. This is due to the fact that some boundary charges do not transform like a primary conformal field:

\begin{eqnarray}
\mathcal{L}(z_1)\mathcal{L}(z_2)&\sim&\frac{\frac{c}{2}}{z_{12}^4}+\frac{2 \mathcal{L}}{z_{12}^2}+\frac{\mathcal{L}'+\frac{6}{c}\mathcal{G}_{A} \bar{\mathcal{G}}_{A}}{z_{12}}\\
\mathcal{L}(z_1)\mathcal{J}^A(z_2)&\sim&0\\
\mathcal{L}(z_1)\mathcal{G}_{A}(z_2)&\sim&\frac{\frac{3}{2} \mathcal{G}_{A}}{ z_{12}^2}+\frac{\mathcal{G}_{A}'-\frac{6}{c}(\sigma^B)^{*}_{BC}\mathcal{J}^B\mathcal{G}_{C}}{z_{12}}\\
\mathcal{L}(z_1)\bar{\mathcal{G}}_{A}(z_2)&\sim&\frac{\frac{3}{2} \mathcal{G}_{A}}{ z_{12}^2}+\frac{\mathcal{G}_{A}'+\frac{6}{c}(\sigma^B)_{BC}\mathcal{J}^B\mathcal{G}_{C}}{z_{12}}\\
\mathcal{J}^A(z_1)\mathcal{J}^B(z_2)&\sim&\frac{\frac{c}{12}}{z_{12}^2}+\,\frac{i \epsilon^{ABC}}{z_{12}} \mathcal{J}^{C}\\
\mathcal{J}^A(z_1)\mathcal{G}_{B}(z_2)&\sim&\frac{\frac{i}{2}(\sigma^B)_{BC}}{z_{12}}\mathcal{G}_{C}\\
\mathcal{J}^A(z_1)\bar{\mathcal{G}}_{B}(z_2)&\sim&\frac{-\frac{i}{2}(\sigma^B)_{BC}}{z_{12}}\bar{\mathcal{G}}_{C}\\
\mathcal{G}_{A}(z_1)\mathcal{G}_{B}(z_2)\,&\sim & \,\delta^{AB}\left({{2c\over 3}\over{z_{12}^{3}}}\,+\,{{2(\mathcal{L}+\frac{6}{c}\mathcal{J}^{C}\mathcal{J}^{C})}\over{z_{12}}}\right)\nonumber\\
&{}&-4(\sigma^C)_{AB}\left({{\mathcal{J}^C}\over{z_{12}^{2}}}\,+\,{{\frac{1}{2}{\mathcal{J}^C}'}\over{z_{12}}}\right)
\end{eqnarray}

Additionally, there are nonlinear terms such as $(\mathcal{J^AJ^A})(z)$, $(\mathcal{G}_{A}\bar{\mathcal{G}}_{A})(z)$, and $(\mathcal{J}^A\mathcal{G}_{B})(z)$, as discussed in the previous section (Eq. (\ref{bhreduction1})). Therefore, it is necessary to consider some redefinitions on the boundary charges and gauge parameters as

\begin{equation}\label{shift1}
\mathcal{L}\rightarrow\mathcal{L}+\frac{3}{2c} \mathcal{J}^A\mathcal{J}^A,\,~~~~\varrho_A\rightarrow\varrho_A-\frac{3}{c}\epsilon\mathcal{J}^A.
\end{equation}

It is crucial to highlight that these novel variables do not influence the boundary charges. This observation ultimately results in operator product expansions of a suitably chosen asymptotic symmetry algebra for $\mathcal{N}=(4,4)$ superconformal boundaries. The set of conformal generators $\mathcal{G}_{A}$ is transformed as $\mathcal{G}_{A}\rightarrow\mathcal{G}_1\pm\mathcal{G}_2$, and $\bar{\mathcal{G}}_{A}\rightarrow\bar{\mathcal{G}}_1\pm\bar{\mathcal{G}}_2$ in the complex coordinates, utilizing Eq. (\ref{Qvar2}). By applying the identical procedure for the barred sector, it can be inferred that the asymptotic symmetry algebra for the most liberal set of boundary conditions of $\mathcal{N}=(4,4)$ supergravity comprises two copies of the $\mathcal{N}=4$ Superconformal algebra with a central charge $c=6k$.
\newpage

\section{Discussions}
In this study, we examined the most general (affine) and superconformal boundary conditions in $AdS_3$  with negative cosmological constant for the $\mathcal{N}=(3,3)$ and $\mathcal{N}=(4,4)$ supergravity theories. In this section, we provide some details on the conservation of boundary charges and offer insights into the topic.

\subsection{Conservation of Boundary Charges}
The boundary conditions for  $\mathcal{N}$ supersymmetries presented in this work lead to non-trivial but integrable canonical boundary charges, thus generating a meaningful asymptotic symmetry algebra. Here, we elaborate on the conservation of boundary charges for the affine cases of $\mathcal{N} = 3$ and $\mathcal{N} = 4$. The boundary conditions presented in this work, given in Eqs. \eqref{flatness} - \eqref{conn}, lead to canonical boundary charges with significant properties for both the $\mathcal{N}=3$  \eqref{boundaryco9} and $\mathcal{N}=4$  \eqref{boundaryco13}affine cases, respectively.

The conservation of these canonical boundary charges is directly related to the specification of the boundary conditions. For the $\mathcal{N} = 3$ affine case, the time evolution of the charge is given by:
\begin{align}
\partial_t \mathcal{Q}[\lambda]& = \frac{k}{2} \int  d\varphi \Big( \epsilon^{+1} \partial_\varphi \mu^{-1} 
- \frac{1}{2} \epsilon^{0} \partial_\varphi \mu^{0} \\
&+ \epsilon^{-1} \partial_\varphi \mu^{+1}
+ \varrho_A \partial_\varphi \eta^{A} \Big) \\
&+ 2 \Big( \zeta_{A}^{-p} \partial_\varphi \nu_{A}^{p}
- \zeta_{A}^{p} \partial_\varphi \nu_{A}^{-p} \Big),
\end{align}
and for the $\mathcal{N}=4$ affine case:
\begin{align}
\partial_t \mathcal{Q}[\lambda] &= \frac{k}{2} \int  d\varphi \Big( \epsilon^{+1} \partial_\varphi \mu^{-1} 
- \frac{1}{2} \epsilon^{0} \partial_\varphi \mu^{0} \\
&+ \epsilon^{-1} \partial_\varphi \mu^{+1}
+ \varrho_A \partial_\varphi \eta^{A} \Big) \\
&+ 2 \Big( \zeta_{A}^{-p} \partial_\varphi \nu_{A}^{p}
- \zeta_{A}^{p} \partial_\varphi \nu_{A}^{-p} \Big)\\
&+ 2 \Big( \bar{\zeta}_{A}^{-p} \partial_\varphi \bar{\nu}_{A}^{p}
-\bar{\zeta}_{A}^{p} \partial_\varphi \bar{\nu}_{A}^{-p} \Big).
\end{align}
The constancy of the chemical potentials plays a crucial role in the temporal conservation of the boundary charges. Specifically, the conservation of boundary charges depends on whether the chemical potentials are independent of the angular coordinate $\varphi$. If the chemical potentials are constant or independent of $\varphi$, then the boundary charge is conserved, i.e., $\partial_t Q[\lambda] = 0.$ However, if the chemical potentials depend on $\varphi$, then temporal conservation of the boundary charges still holds, but this leads to $\partial_t \delta Q[\lambda] = 0.$ In this case, variations in the chemical potentials are not allowed, thus establishing a conservation law.
\subsection{Insights into $\mathcal{N} > 2$ topic}

$\mathcal{N} = 3$ and $\mathcal{N} = 4$ Chern\,-\,Simons theories significantly clarify aspects that remain obscure in the $\mathcal{N} = 1$ and $\mathcal{N} = 2$ cases. Specifically, the supersymmetric structures in $\mathcal{N} = 3$ and $\mathcal{N} = 4$ allow for the emergence of more complex symmetry structures, providing a deeper understanding of conserved charges and interactions. In these theories, $AdS/CFT$ duality becomes more pronounced, playing a critical role in understanding the relationship between asymptotic symmetry algebras and the $\mathcal{N} = (4,4)$  supersymmetric structures in boundary conformal field theories.
This study not only focuses on $AdS_3$ but also highlights the role of the $AdS_3 \times S_3 \times T^4$ background from the perspective of string theory. Furthermore, the primary motivation for examining cases where
$\mathcal{N} > 2$ arises from their connection to supersymmetric string compactifications. For instance, the $AdS_3 \times S_3 \times T^4$ background provides $\mathcal{N} = (4,4)$ supersymmetry. Moreover, there is evidence suggesting that higher-spin supergravity can be consistently embedded into string theory; for concrete examples, see \cite{Gaberdiel:2014cha} and \cite{Eberhardt:2018ouy}. Finally, analyses of the moduli space and different flux configurations enable a more comprehensive investigation of the underlying dynamics. In this context, the study explores the potential of Chern\,-\,Simons theory and its relationship to string theory, demonstrating that the asymptotic symmetry algebra in $\mathcal{N} = (4,4)$ Chern\,-\,Simons theory can be interpreted as the $\mathcal{N} = (4,4)$ space-time symmetry in string theory.

\section{Concluding Remarks}
\hspace{0.5cm}
In conclusion, our study has examined the comprehensive examples of the  $\mathcal{N}=(3,3)$ and $\mathcal{N}=(4,4)$ formulations of $\mathfrak{osp}(N|2)$ Chern-Simons supergravity on $AdS_3$, 
inspired by the $\mathcal{N}<3$ cases studied previously\cite{Ozer:2021wtx,Ozer:2019nkv}.\,The focus on  $\mathcal{N}=(3,3)$ and $\mathcal{N}=(4,4)$ cases has been instrumental, shedding light on the asymptotic structure of these extended supergravity theories. Our findings reveal that, for the loosest set of boundary conditions, the asymptotic symmetry algebras of these formulations are two copies of the  $\mathfrak{osp}(3|2)_k$ and $\mathfrak{osp}(4|2)_k$ affine algebras, respectively. Additionally, by constraining the connection fields on the boundary conditions, we successfully obtained the supersymmetric extensions of the Brown-Henneaux boundary conditions.

Building upon these results, we have conclusively identified that the asymptotic symmetry algebras for  $\mathcal{N}=(3,3)$ and $\mathcal{N}=(4,4)$ extended higher-spin supergravity theories in$AdS_3$ are two copies of the $\mathcal{N}=3$ and $\mathcal{N}=4$  superconformal algebras, respectively. This significant contribution enhances our understanding of the asymptotic structure of these extended supergravity theories and provides a valuable foundation for future researchers exploring AdS supergravity models.

The $\mathcal{N}$ supersymmetry boundary conditions presented in this work lead to integrable canonical boundary charges, generating a meaningful asymptotic symmetry algebra. Specifically, the conservation of boundary charges for both $\mathcal{N} = 3$ and $\mathcal{N} = 4$ affine cases has been clearly demonstrated. The constancy of the chemical potentials plays a decisive role in the conservation of boundary charges. This work explicitly shows that the chemical potentials are independent of the angular coordinate $\varphi$, ensuring that the boundary charges are conserved. Consequently, the boundary conditions established in this work provide a solid foundation for preserving the asymptotic symmetry algebra.

Furthermore, our study hints at a potential relationship between higher spin theory and string theory, as suggested in our prior work on $\mathcal{N}=2$.\,While we briefly touched upon this connection, we emphasize the importance of examining the physical implications of integrating higher spin theories into the broader framework of string theory, particularly for theories with reduced supersymmetry. Our illustrative example involving the $\mathcal{N}=4$  higher spin theory in the limit $\lambda \rightarrow 0$ within superstring theory on $AdS_3 \times S^3 \times T^4$ serves as a starting point.\,The groundwork for this understanding was initially laid out in \cite{Gaberdiel:2014cha}, and more recently, the relevant string theory has been meticulously developed, as documented in \cite{Gaberdiel:2018rqv,Eberhardt:2018ouy} and \cite{Eberhardt:2019ywk}.

However, we acknowledge that a comprehensive exploration of such relationships requires further investigation, posing questions like "Can we pinpoint the higher spin theory corresponding to the theory presented in \cite{Datta:2017ert}?" and "Can we discern the string theory that aligns with higher spin theories?" This exploration holds the potential to uncover specific junctures where our higher spin theory may converge to a limit, building upon the foundation laid in the previous study and leaving room for future discoveries.\,Our conclusions open avenues for future research, offering insights into the interplay between higher spin theories and string theory within AdS supergravity.

As a result, this work endeavors to deepen our understanding of three-dimensional supersymmetry and fields with spin greater than two in Einstein's supergravity theories, while emphasizing their connection with string theory. The primary objective is to comprehensively develop these theories and explicitly determine their asymptotic symmetries under the most general boundary conditions, as outlined in the introduction.

One focal point of the research is understanding the asymptotic behavior of established supersymmetric theories and identifying their asymptotic symmetries under the most general boundary conditions. Asymptotic symmetries play a crucial role in comprehending theories' behavior at large distances, offering significant insights into their general behavior.
Additionally, the work highlights the process of associating the results of the developed Einstein supergravity theories with string theory. However, detailed discussions on the straightforward aspects of this relationship are deferred to a subsequent study.\\

By deferring the straightforward details of the relationship with string theory to a future work, this work aims to pave the way for further exploration. The intention is to provide a comprehensive foundation for the mathematical structure of supersymmetric theories, leaving the explicit exploration of the intricate details of the interplay between these theories and string theory for subsequent research endeavors.\\
 Finally, in this study, we analyzed the boundary conditions for Chern--Simons theories based on the superalgebras $\mathfrak{osp}(3|2)$ and $\mathfrak{osp}(4|2)$ , corresponding to  $\mathcal{N}=(3,3)$ and $\mathcal{N}=(4,4)$ supersymmetric boundary $SCFTs$, respectively. However, the large $\mathcal{N}=(4,4)$ supersymmetric theory, which{\color{black}{ is dual to $AdS_3 \times S^3\times S^3 \times S^1$}}, was not addressed in this paper due to the additional challenges introduced by the extended geometric and symmetry structures. This significant topic, requiring a more detailed analysis, is reserved for future work.\\

Additionally, our approach to deriving the $OPEs$ of the boundary $SCFT$ currents is closely related to the methodology outlined in \cite{Creutzig:2012xb} and \cite{Gutperle:2011kf}, where $OPEs$ are derived from the bulk Chern--Simons equations of motion. However, unlike these studies, our approach emphasizes the role of supersymmetry in shaping the $\cW-$ algebra structures on the boundary and constructs these structures explicitly from the underlying supergroups. Therefore, our work provides a complementary perspective, contributing to a deeper understanding of the interplay between bulk symmetries and boundary dynamics.
{\color{black}{While the methods in \cite{Creutzig:2012xb} and \cite{Gutperle:2011kf} are significant, they have not been extended to supersymmetric cases such as the ones discussed here. Thus, our approach generalizes these methods, offering a unique contribution by focusing on the detailed investigation of supersymmetric structures.
}}
\section{Acknowledgments}
\textit{
The authors are thankful to Jan de Boer, Matthias R. Gaberdiel,Carlos E.Valc\'arcel Flores , Changhyun Ahn  and Lorenz Eberhardt for interest to this work and useful comments.\\
{\color{black}{The authors are supported by The Scientific and Technological Research Council of T\"urkiye (T\"UB\'ITAK) through the ARDEB 1001 project with Grant number 123F255.}} 
}


\end{document}